\documentclass[10pt]{report}

\usepackage[utf8]{inputenc}
\usepackage[T1]{fontenc}
\usepackage[english]{babel}
\usepackage{amsthm,amsmath,amssymb}
\usepackage{color}
\usepackage{enumitem}
\usepackage{framed}
\usepackage{graphicx}
\usepackage{caption}
\usepackage{subcaption}
\usepackage{pdfpages}
\usepackage{url}
\usepackage[left=4.5cm,right=4.5cm]{geometry}

\newtheorem{theorem}{Theorem}

\newtheorem{lemma}{Lemma}

\newtheorem{definition}{Definition}
\newtheorem{remark}{Remark}

\DeclareMathOperator{\mat}{Mat}
\DeclareMathOperator{\rank}{rk}
\DeclareMathOperator{\nrm}{Nm}
\DeclareMathOperator{\Nrm}{N}
\DeclareMathOperator{\rnm}{nm}
\DeclareMathOperator{\tra}{Tr}
\DeclareMathOperator{\rtr}{tr}
\DeclareMathOperator{\diag}{diag}

\DeclareMathOperator*{\argmin}{arg\,min}
\DeclareMathOperator*{\argmax}{arg\,max}
\DeclareMathOperator{\snr}{SNR}
\DeclareMathOperator{\vol}{vol}
\DeclareMathOperator{\Hom}{Hom}

\DeclareMathOperator{\disc}{disc}
\DeclareMathOperator{\Cl}{Cl}
\DeclareMathOperator{\ind}{ind}
\DeclareMathOperator{\info}{I}
\DeclareMathOperator{\gal}{Gal}

\newcommand{\Q}{\mathbb{Q}}

\newcommand{\Z}{\mathbb{Z}}
\newcommand{\R}{\mathbb{R}}
\newcommand{\C}{\mathbb{C}}
\newcommand{\F}{\mathbb{F}}

\newcommand{\mb}[1]{\mathbf{#1}}
\newcommand{\mc}[1]{\mathcal{#1}}
\newcommand{\mf}[1]{\mathfrak{#1}}
\newcommand{\vl}[1]{\vol\left(#1\right)}
\newcommand{\rk}[1]{\rank\left(#1\right)}
\newcommand{\nm}[2]{\nrm_{#1}\left(#2\right)}
\newcommand{\Nm}[2]{\Nrm_{#1}\left(#2\right)}
\newcommand{\tr}[2]{\tra_{#1}\left(#2\right)}
\newcommand{\dm}[1]{\dim\left(#1\right)}

\usepackage{tikz}
\usetikzlibrary{arrows, automata, positioning, decorations.pathreplacing, decorations.markings, shapes.geometric}
\tikzset{
blank/.style={minimum height=2em,minimum width=3cm,text centered},
blank_short/.style={minimum height=2em,minimum width=1cm,text centered},
user/.style={rectangle,rounded corners,draw,minimum height=2em,minimum width=1cm,text centered},
dummy/.style={minimum height=2em,minimum width=1.5cm,text centered},
relay/.style={rectangle,rounded corners,draw,minimum height=1em,minimum width=1cm,text centered},
pil/.style={->,dashed,shorten <=2pt,shorten >=2pt},
pil_rev/.style={<-,shorten <=2pt,shorten >=2pt},
radiation/.style={{decorate,decoration={expanding waves,angle=90,segment length=4pt}}},
solidnode/.style={rectangle, draw=black, minimum width=1cm, minimum height=0.75cm},
dashednode/.style={rectangle, draw=black, dashed, minimum width=1cm, minimum height=0.75cm},
}

\begin{document}

\begin{titlepage}
\centering

\LARGE{Lattice Codes for Physical Layer Communications}\par\bigskip
\bigskip
\Large{Amaro Barreal Fern\'andez}\par\bigskip
\bigskip
\large{Department of Mathematics and Systems Analysis \\ Aalto University \\ Finland} \par\bigskip
\bigskip
\bigskip
Doctoral Dissertation \par\bigskip

\flushleft\vspace{\fill}
\textbf{Supervising Professor:}\par  
Prof. Camilla Hollanti\par 
Aalto University\par 
Finland
\end{titlepage}

\chapter*{List of Publications}
This thesis consists of an overview and the following publications which are referred to in the text by their Roman numerals. \par 
\bigskip
\bigskip

\noindent \textbf{Publication I:}\par 
\noindent \textit{Natural Orders for Asymmetric Space--Time Coding: Minimizing the Discriminant.}\par 
\noindent Amaro Barreal, Capi Corrales Rodrig\'a\~nez, and Camilla Hollanti.\par 
\noindent Submitted, 2016. \par\bigskip\medskip

\noindent \textbf{Publication II:}\par  
\noindent \textit{Constructions of Fast-Decodable Distributed Space--Time Codes}\par 
\noindent Amaro Barreal, Camilla Hollanti, and Nadya Markin \par 
\noindent CIM Series in Mathematical Sciences, Springer, 3, pp. 43--51, 2014.\par\bigskip\medskip

\noindent \textbf{Publication III:} \par
\noindent \textit{Fast-Decodable Space--Time Codes for the N-Relay and Multiple-Access MIMO Channel} \par
\noindent Amaro Barreal, Camilla Hollanti, and Nadya Markin\par 
\noindent IEEE Transactions on Wireless Communications, 15(3), pp. 1754--1767, 2015.\par\bigskip\medskip

\noindent \textbf{Publication IV:} \par
\noindent \textit{Theta Series Approximation with Applications to Compute-and-Forward Relaying} \par
\noindent Amaro Barreal, David Karpuk, and Camilla Hollanti\par 
\noindent Revision submitted, 2017.\par\bigskip\medskip

\noindent \textbf{Publication V:} \par
\noindent \textit{A Low-Complexity Message Recovery Method for Compute-and-Forward Relaying} \par
\noindent Amaro Barreal, Joonas Pääkkönen, David Karpuk, Camilla Hollanti, and Olav Tirkkonen\par 
\noindent IEEE Information Theory Workshop - Fall, pp. 39--43, 2015.\newpage
\noindent \textbf{Publication VI:} \par
\noindent \textit{Well-Rounded Lattices for Coset Coding in MIMO Wiretap Channels} \par
\noindent Oliver Gnilke, Amaro Barreal, Alex Karrila, Ha Tran, David Karpuk, and Camilla Hollanti\par 
\noindent IEEE International Telecommunication Networks and Applications Conference, pp. 289--294, 2016.\par\bigskip\medskip

\noindent \textbf{Publication VII:} \par
\noindent \textit{Information Bounds and Flatness Factor Approximation for Fading Wiretap MIMO Channels} \par
\noindent Amaro Barreal, Alex Karrila, David Karpuk, and Camilla Hollanti\par 
\noindent IEEE International Telecommunication Networks and Applications Conference, pp. 277--282, 2016.

\chapter{Introduction}
\label{chp:intro}
It is a matter of course in our society that wireless data exchange can easily be accomplished fast, reliably and securely. However, the challenges faced for wireless networks to live up to these expectations are daunting. Additional devices are introduced daily to wireless networks, increasing not only the density of the system, but also acting as additional potential sources of interference, a destructive phenomenon that is difficult to handle, and often results in unsuccessful data exchange. Communication considering the actual physical transmission medium is said to happen at the \emph{physical layer} \cite{osi}. 

In recent years, data traffic worldwide has reached incredible numbers. The amount of data carried by mobile networks each month is in the order of exabytes, growing 18-fold over the last five years. Especially mobile video traffic has recently seen unprecedented growth, and the numbers will keep rising as millions of different mobile devices and connections are added annually around the globe. Due to the forced heterogeneity of our networks, which need to accommodate many new types of devices, reliable codes for traditional downlink communications are often not suitable for serving modern transmission protocols. More modern communication schemes, derived over the last decade, have the potential to become indispensable in future wireless generations. For example, the future $5^\mathrm{th}$ Generation (5G) wireless systems will incorporate many different techniques, including distributed antenna systems and massive multiple-input multiple-output systems. The mentioned incompatibility is mainly due to devices involved in the transmission process being equipped with unequal numbers of antennas or differing in available computational power. Such incompatibility  can also be a result of enabling security directly at the physical layer, increasing the throughput of the system, etc. This motivates the study of well-performing codes for many of those novel transmission schemes, and is the main catalyst behind the research leading to this thesis. There are however many aspects and characteristics offered by traditional downlink codes that remain beneficial for modern approaches. The mathematical link is the underlying lattice structure, an omnipresent object in the construction of physical layer codes. The study of lattice codes has often provided additional motivation for studying purely mathematical problems, which are interesting in their own right. Though lattices are highly symmetric and regular objects, their simple structure is deceptive, and many problems in lattice theory remain open.    

Going back to 1983, Conway and Sloane introduced a simple encoding method based on the concept of a Voronoi code \cite{conway:voronoi}, for which codewords are essentially the collection of coset representatives of a quotient of lattices. Over the following years, several criteria have been developed for constructing Voronoi codes which offer reliable performance. More importantly, many significant existence results related to sequences of lattices achieving certain asymptotic characteristics have been proven by Poltyrev~\cite{poltyrev:nested}, Zamir and Feder~\cite{zamir:nested}, and many others, additionally motivated by practical applications. Even before the concept of Voronoi codes was introduced, the seminal 1975 paper by Wyner introduced the so-called wiretap channel, though at the time in a wired setting \cite{wyner:wiretap}. Both of these concepts have come together in a modern reinterpretation of the wiretap channel, now in a wireless context. Moreover, the recent award-winning compute-and-forward protocol developed by Nazer and Gastpar \cite{nazer:cf} relies on the use of Voronoi codes for achieving high computation rates. This protocol has received much deserved attention since its introduction in 2008, and is considered one of the most relevant modern relaying schemes. 

Similarly but in the context of multiple-antenna downlink communications stand the so-called space--time codes. The first construction of a space--time code goes back to 1998, the famous Alamouti code for a system with two transmit antennas \cite{alamouti:stc}. Even if the construction of this particular code was achieved from an exclusively engineering perspective and mathematically unmotivated, it was noticed later that codewords correspond precisely to the left-regular representation of elements of the Hamilton quaternions $\mathbb{H}$. The discovery of the Alamouti code was the starting point of a long and fruitful line of research involving both engineers and mathematicians. Eventually, division algebras were proposed to serve as underlying structures by Sethuraman et al. \cite{sethuraman:stc}, which led to the construction of multiple extraordinary codes, such as the Golden code \cite{belfiore:golden}, which was later incorporated in the IEEE 802.16 (WiMAX) standard, or general Perfect codes \cite{oggier:perfect}. Furthermore, the usefulness of maximal orders within the considered algebras was then discovered by Hollanti and Lahtonen \cite{hollanti:order1}, which further improved the potential performance of codes arising from these structures, though at the cost potentially difficult bit labeling. 

The high complexity of lattice decoders has traditionally been the bottleneck for practical implementation of optimal decoders for space--time lattice codes. While it is possible to resort to suboptimal decoders, the meticulously ensured good performance of space--time codes would suffer. A potential algebraic decoding complexity reduction was first addressed in \cite{biglieri:fd} and motivated further related work, giving rise to different families of so-called fast-decodable space--time codes. Even though it has been recently shown by Mejri et al. \cite{mejri:fd_revisited} that the usual approaches do not capture all families of fast-decodable codes, the typical methods allow for explicit algebraic conditions which enable fast-decodability. This term can be rather misleading, as the decoding complexity of fast-decodable codes can often still lead to very slow decoding. However, fast-decodable codes offer a reduction in decoding complexity in contrast to non-fast-decodable codes with comparable properties.   

Though neglected in recent years, the asymmetry found in modern networks and contemporary communication protocols naturally enable the use of more specialized space--time codes. For instance, many recently developed physical layer relaying protocols, such as the multiple-antenna amplify-and-forward scheme \cite{nabar:relay,yang:af} from 2007, relies on cleverly constructed space--time codes. Thus, space--time codes are again becoming more relevant for modern wireless communications. Furthermore, space--time codes often arise by force of nature in certain communication setups, such as the multiple-access channel. It is hence not always about choice and design only. In its broad generality, the study of lattice codes for physical layer communications is an interesting and rapidly evolving area of multidisciplinary research. 

This thesis is composed of multiple articles in this interdisciplinary area of research, and we give a brief overview on how all considered settings and results are linked together. This summary is structured as follows. We recall some of the most important mathematical objects and results related to lattice code design in Chapter~\ref{chp:math}. Therein, we start with concepts from algebraic number theory, followed by the theory of central simple algebras and their orders. Finally, we study lattices and their properties, the most important mathematical object in this thesis.  
We follow up with Chapter \ref{chp:plc}, where we introduce the basic principles and characteristics of wireless communications on the basis of a simple point-to-point channel. We furthermore introduce the notion of space--time codes and nested lattice codes in Section~\ref{subsec:stc} and \ref{subsec:nested_codes}, respectively. 
The results of the publications composing this thesis are then discussed in Chapter~\ref{chp:communications}. We divide the chapter into four sections, each corresponding to a different wireless communications setting. In each of the sections, the considered communications protocol is explained in detail, and the goals and results of the corresponding publications are put in context. 

\chapter{Mathematical Preliminaries}
\label{chp:math}

In this chapter, we acquaint the reader with some of the mathematical notions in algebraic number theory, class field theory, and the theory of lattices that are most important to this thesis. As a main reference for the number theoretic results presented in Sections~\ref{sec:ant} and \ref{sec:csa} we refer to \cite{milne:cft, milne:ant, neukirch:ant}, while \cite{conway:lattices, ebeling:lattices} serve as references for all lattice related concepts introduced in Section~\ref{sec:lattices}.  

\section{Algebraic Number Theory}
\label{sec:ant}

We begin with the notion of algebraic number fields. Let $L/K$ be an arbitrary field extension. An element $\alpha \in L$ is called \emph{algebraic} over $K$ if there exists a non-zero polynomial $f(x) \in K\left[x\right]$ such that $f(\alpha) = 0$, and the field extension $L/K$ is called \emph{algebraic} if all elements of $L$ are algebraic over $K$. Every field extension of finite degree $\left[L:K\right] := \dim_{K}(L) < \infty$ is algebraic. 

\begin{definition}
	An \emph{algebraic number field} is a finite extension of $\Q$. 
\end{definition}

To every number field $K$ we can associate its \emph{ring of integers} $\mc{O}_K$, the unique integral closure of $\Z$ in $K$. In other words, $\mc{O}_K$ is the collection of all elements of $K$ which satisfy a monic polynomial equation with coefficients in $\Z$. In the simplest case $K = \Q$, we have $\mc{O}_K = \Z$, and we see that $K$ is the field of fractions of $\mc{O}_K$. This statement remains true for all number fields.

\subsection{Norm, Trace and Discriminant}
\label{subsec:discriminant}

Given a finite number field extension $L/K$ of degree $n$, every $\alpha \in L$ naturally defines a $K$-linear endomorphism $L \to L;\ l \mapsto \alpha l$ with well-defined norm and trace, which we refer to as the \emph{relative norm} $\nm{L/K}{\alpha}$ and \emph{relative trace} $\tr{L/K}{\alpha}$ of the field extension, respectively. We fix compatible embeddings of $K$ and $L$ into $\C$, and identify the fields with their images under these embeddings. More precisely, there exist exactly $n$ pairwise distinct embeddings $\sigma_i:L \to \C$, such that $\sigma_i\mid_K$ is the identity on $K$. Let $\Hom_K(L,\C) = \left\{\sigma_1,\ldots,\sigma_n\right\}$. We have for all $\alpha \in L$
\begin{align*}
	\nm{L/K}{\alpha} = \prod\limits_{i=1}^{n}{\sigma_i(\alpha)}; \qquad \tr{L/K}{\alpha} = \sum\limits_{i=1}^{n}{\sigma_i(\alpha)}.
\end{align*}

The trace form can be used to define an important invariant of a number field. In fact, viewing $L$ as a $K$-vector space with basis $\left\{b_1,\ldots,b_n\right\}$, the trace 
\begin{align*}
	\tra_{L/K}: L\times L \to K; \quad (\alpha,\beta) \mapsto \tr{L/K}{\alpha\beta}
\end{align*}
is a non-degenerate symmetric bilinear form on $L$, with corresponding matrix $T(b_1,\ldots,b_n) = \left(\tr{L/K}{b_i b_j}\right)_{i,j}$. Using the description of the trace form in terms of $\Hom_K(L,\C)$, we have $\det(T(b_1,\ldots,b_n)) = \det\left(\sigma_i(b_j)\right)_{i,j}^2$. If $\left\{b_1,\ldots,b_n\right\}$ is an integral basis, the ideal $\disc(b_1,\ldots,b_n) = \langle \det\left(\sigma_i(b_j)\right)_{i,j}^2 \rangle$ is called the \emph{relative discriminant} $\disc(L/K)$ of the extension $L/K$.

Dedekind showed the existence of an integral basis for every number field extension $L/K$. When $\mc{O}_K$ is a principal ideal domain, then every finitely generated $\mc{O}_L$-module is free over $\mc{O}_K$, and we even have an integral basis of $\mc{O}_L$. In this case, the discriminant of the extension $L/K$ can be seen as an integer rather than an ideal, as it is independent of the basis up to a unit factor. In particular, if $K/\Q$ is an extension of degree $n$, the ring of integers $\mc{O}_K$ is finitely generated as a $\Z$-module of rank $n$. We have the following definition. 

\begin{definition}
	Let $K$ be a number field of degree $n$, with ring of integers $\mc{O}_K$, and let $\left\{b_1,\ldots,b_n\right\}$ be an integral basis of $\mc{O}_K$. The \emph{discriminant} of $K$ is the well-defined integer $d_K = \disc(b_1,\ldots,b_n) = \disc(\mc{O}_K/\Z)$. 
\end{definition}

\subsection{The Group of Units}
\label{subsec:units}

An important subset of the ring of integers $\mc{O}_K$ is its group of units, $\mc{O}_K^\times$. Let us divide the group $\Hom_\Q(K,\C)$ into the subset of real embeddings $\left\{\sigma_1,\ldots,\sigma_r\right\}: K \to \R$, and the subset of pairs of complex embeddings $\left\{\sigma_{r+1},\overline{\sigma}_{r+1},\ldots,\sigma_s,\overline{\sigma}_s\right\}: K \to \C$. We have $n = r+2s$, and call $(r,s)$ the \emph{signature} of the number field $K$.

An element $\alpha \in \mc{O}_K$ is a unit if and only if $\nm{K/\Q}{\alpha} = \pm 1$, so that 
\begin{align*}
	\prod\limits_{i=1}^{r}{|\sigma_i(\alpha)|}\prod\limits_{j=r+1}^{r+s}{|\sigma_j(\alpha)|^2} = 1.
\end{align*} 

If we denote by $\mu\left(\mc{O}_K^\times\right)$ the roots of unity, a result due to Dirichlet tells us that $\mc{O}_K^\times$ is a finitely generated abelian group of rank $r+s-1$, and as such, 
%
%
%
	\begin{align*}
		\mc{O}_K^\times = \mu\left(\mc{O}_K^\times\right) \oplus \Z^{r+s-1}.
	\end{align*}

For all real number fields, that is, of signature $(r,0)$, we have $\mu\left(\mc{O}_K^\times\right) = \left\{\pm 1 \right\}$. Moreover, note that for imaginary quadratic number fields we have $(r,s) = (0,1)$, and the rank of $\mc{O}_K^\times$ is thus zero. This implies a finite group of units, and it is the only case where this occurs, excluding the trivial case $K = \Q$. Otherwise, the group of units has infinite cardinality.

\subsection{Ideals and Ramification}
\label{subsec:ramification}

The motivation for studying number fields has its origins in the factorization of integers into primes. In the ring $\Z$, prime and irreducible elements coincide, and as we know every natural number factors uniquely into prime numbers. By generalizing the ring $\Z$ to the ring of integers $\mc{O}_K$ of a number field, unique factorization into prime elements is no longer guaranteed. However, the underlying structure of the ring $\mc{O}_K$ allows for a generalization of unique factorization by making use of ideals, instead of elements. 

Let $L/K$ be a number field extension of degree $n$. Any prime ideal $\mf{p}$ of $\mc{O}_K$ factors in $\mc{O}_L$ as
\begin{align*}
	\mf{p}\mc{O}_L = \prod\limits_{i=1}^{g}{\mf{P}_i^{e_i}},
\end{align*} 
where $\mf{P}_i$ are non-zero distinct prime ideals of $\mc{O}_L$, and $e_i > 0$. We say that the primes $\mf{P}_i$ \emph{lie} over $\mf{p}$. If one of the exponents satisfies $e_i > 1$, we say that $\mf{p}$ is \emph{ramified} in $\mc{O}_L$, and refer to the number $e_i = e(\mf{P}_i/\mf{p})$ as the \emph{ramification index}. We further define the \emph{residue class degree} $f_i = \left[\mc{O}_L/\mf{P}:\mc{O}_K/\mf{p}\right]$. On the other hand, if $e_i = f_i = 1$ for all $i$, then $\mf{p}$ \emph{splits completely} in $L$, and we say that it is \emph{inert} in $L$ if $\mf{p}$ remains prime in $\mc{O}_L$. 

The ramification indices and residue class degrees are related via the important identity 
\begin{align*}
	\sum\limits_{i=1}^{g}{e_i f_i} = n,
\end{align*}
which simplifies to $efg = n$ for Galois extensions. 

An extension $L$ of a number field $K$ is \emph{unramified} over $K$ if no prime ideal of $\mc{O}_K$ ramifies in $\mc{O}_L$. While there do not exist unramified extensions of $\Q$, there may exist unramified extensions of other number fields. A prominent example is the maximal unramified abelian extension of a number field $K$, known as the \emph{Hilbert class field} of $K$. 

We can define the relative norm of an ideal $\mf{P} \subset \mc{O}_L$ lying over $\mf{p} \subset \mc{O}_K$ by defining a homomorphism $\Nrm_{L/K}$ from the set of ideals of $\mc{O}_L$ to the set of ideals of $\mc{O}_K$ as 
\begin{align*}
	\Nm{L/K}{\mf{P}} := \mf{p}^{f(\mf{P}/\mf{p})},
\end{align*}
where $\mf{p} = \mf{P}\cap \mc{O}_K$. We simply write $\Nrm(\mf{p})$ when the extension is $K/\Q$. With this definition, for any non-zero ideal $\mf{p} \subset \mc{O}_K$ we have $\Nm{L/K}{\mf{p}\mc{O}_L} = \mf{p}^n$.

The notion of an ideal can be slightly generalized, giving rise to one of the most important invariants of a number field. A \emph{fractional ideal} of $\mc{O}_K$ is a finitely generated non-zero $\mc{O}_K$-submodule $\mf{f}$ of $K$. The collection $J_K$ of fractional ideals forms an abelian group, called the ideal group of $K$. The \emph{ideal class group} $\Cl_K$ of $K$ is the quotient 
\begin{align*}
	\Cl_K := J_K/P_K,
\end{align*}
where $P_K$ is the subgroup of principal fractional ideals. We have the following important result.

\begin{theorem}
	The class number $h_K := |\Cl_K|$ of a number field $K$ is finite. 
\end{theorem}

From the definition we note that the ring of integers $\mc{O}_K$ is a principal ideal domain if and only if $h_K = 1$.

We move on to the local case. Let $K$ be a number field, and $\nu: K \to \R \cup \left\{\infty \right\}$ a valuation on $K$. To a pair $(K,\nu)$ we can assign an absolute value $|\cdot|_{\nu}: K \to \R$, which for $\alpha,\beta \in K$ satisfies the properties $|\alpha|_{\nu} \ge 0$ with $|\alpha|_{\nu} = 0 \Leftrightarrow \alpha = 0$, $|\alpha\beta|_{\nu} = |\alpha|_{\nu}|\beta|_{\nu}$, and $|\alpha+\beta|_{\nu} \le |\alpha|_{\nu}+|\beta|_{\nu}$. The absolute value is called \emph{non-archimedean} if it satisfies the stronger property $|\alpha+\beta|_{\nu} \le \max\left\{|\alpha|_{\nu},|\beta|_{\nu}\right\}$, and is otherwise called \emph{archimedean}.

Two absolute values are said to be \emph{equivalent} if they induce the same topology on $K$, and we call an equivalence class of absolute values on $K$ a \emph{place} of $K$. There exists exactly one place of $K$
\begin{itemize}
	\item[i)] for each prime ideal $\mf{p} \subset \mc{O}_K$, given by 
			$|\alpha|_\mf{p} = (1/\Nrm(\mf{p}))^{\nu_{\mf{p}}(\alpha)}$.
		These places are called \emph{finite places}. 
	\item[ii)] for each real embedding $\sigma:K \to \R$, given by 
			$|\alpha|_\sigma = |\sigma(\alpha)|$,
		and referred to as \emph{infinite real places}.
	\item[iii)] for each conjugate pair of complex embeddings $\sigma:K \to \C$, 
			$|\alpha|_\sigma = |\sigma(\alpha)|^2$.
		These are called \emph{infinite complex places}.
\end{itemize}

Consider a finite number field extension $L/K$, and let $\mf{v}$ and $\mf{w}$ be a place of $K$ and $L$, respectively. If $|\cdot|_\mf{w}$ restricted to $K$ is equivalent to $|\cdot|_\mf{v}$, we say that $\mf{w}$ \emph{lies} over $\mf{v}$ and write $\mf{w}|\mf{v}$. Every place of $K$ extends to a finite number of places of $L$. For finite places, the concept of ramification agrees with the usual ideal theoretic concept mentioned above, while a real place is ramified if a place above it is complex, and is otherwise unramified. 

We are interested in the completion $K_{\mf{v}}$ of $K$ with respect to a place $\mf{v}$. There are two possibilities. 
\begin{itemize}
	\item[i)] If $\mf{v}$ is an infinite place, then $K_{\mf{v}}$ is isomorphic to $\R$ or $\C$, depending on whether $\mf{v}$ is real or complex, and $|\cdot|_{\mf{v}}$ is equivalent to the standard absolute value. 
	
	\item[ii)] If $\mf{v}$ is a finite place, then $K_{\mf{v}}$ is a finite extension of the $p$-adic field $\Q_p$, and the absolute value is isomorphic to the unique extension of the $p$-adic absolute value. 
\end{itemize}

For our purposes it suffices to only consider the local field $K_\mf{p}$, where $\mf{p} \subset \mc{O}_K$ is a non-zero prime ideal, that is, the completion of $K$ with respect to the valuation $\nu_{\mf{p}}$. If we denote by $p = \chi(\mc{O}_K/\mf{p})$ the characteristic of the residue field, then $K_{\mf{p}}$ is a finite algebraic extension of $\Q_p$. Similarly to the ring of integers $\mc{O}_K$ of $K$, we can naturally assign to $K_{\mf{p}}$ the local ring $\mc{O}_{K_{\mf{p}}}$ of elements of absolute value $\le 1$, with unique prime ideal $\mf{p}\mc{O}_{K_{\mf{p}}}$.

Let $L/K_{\mf{p}}$ be a finite algebraic extension, and let $\mf{q}$ be the unique prime ideal of $\mc{O}_L$.  Then, if $e(\mf{q}|\mf{p}) > 1$ we say that $\mf{p}$ -- or the extension $L/K_{\mf{p}}$ -- is \emph{wildly ramified} if $p \mid e(\mf{q}|\mf{p})$, while it is \emph{tamely ramified} if $\gcd(p,e(\mf{q}|\mf{p})) = 1$. As usual, if $e(\mf{q}|\mf{p}) = 1$ then $\mf{p}$ is \emph{unramified}.

\section{Central Simple Algebras}
\label{sec:csa}

Similarly to the transition from the field $\Q$ to a degree-$n$ extension $K$, we can extend the number field $K$ to an algebra of dimension $n$ over $K$. 

Let $K$ be a field, and $\mc{A}$ a finite-dimensional associative $K$-algebra, not necessarily commutative. If $\mc{A}$ has no non-trivial two-sided ideals, it is called \emph{simple}, and it is \emph{central} if its center is precisely $K$. The algebra $\mc{A}$ is a \emph{division algebra}, or also a \emph{skew field}, if all its non-zero elements are invertible. Every simple $K$-algebra is isomorphic to $\mat(n,D)$ for some $n$ and some division $K$-algebra $D$, unique up to isomorphism. We denote by $\ind(\mc{A}) = \sqrt{\left[D:K\right]}$ the \emph{index}, and by $\deg(\mc{A}) = \sqrt{\left[\mc{A}:K\right]}$ the \emph{degree} of the algebra, and $\mc{A}$ is division if and only if $\ind(\mc{A}) = \deg(\mc{A})$. 

It turns out that when we restrict ourselves to number fields, every $K$-central simple algebra is \emph{cyclic}, and vice versa. More concretely, consider a degree-$n$ cyclic Galois extension $L/K$ of number fields, and denote by $\langle \sigma \rangle = \gal\left(L/K\right)$ its cyclic Galois group. A \emph{cyclic algebra} is a tuple 
\begin{align*}
	\mc{C} = (L/K,\sigma,\gamma) := \bigoplus\limits_{i=0}^{n-1}{e^i L},
\end{align*}
where $e^n = \gamma \in K^\times$ and multiplication satisfies $le = e\sigma(l)$ for all $l \in L$.

\subsection{Representation of Cyclic Division Algebras}
\label{subsec:representation}

As above, fix a degree-$n$ cyclic Galois extension $L/K$ of number fields and a $K$-central cyclic algebra $\mc{C}$ of dimension $n$. We treat $\mc{C}$ as a right $L$-vector space, and fix the basis $\left\{1,e,\ldots,e^{n-1}\right\}$. For fixed $x = \sum_{i=0}^{n-1}{e^i x_i} \in \mathcal{C}$ and all $y \in \mc{C}$, the right $L$-linear map $\rho: y \mapsto xy$ describes left multiplication by elements in $\mc{C}$, and is compatible with algebra multiplication. The associated matrix is given by

\begin{align*}
	x \mapsto \rho(x) := \begin{bmatrix} 
				x_0 & \gamma\sigma(x_{n-1}) & \gamma\sigma^2(x_{n-2}) & \cdots & \gamma\sigma^{n-1}(x_1) \\ 
				x_1 & \sigma(x_0) & \gamma\sigma^2(x_{n-1}) & & \gamma\sigma^{n-1}(x_2) \\
				\vdots & & \vdots & & \vdots \\
				x_{n-2} & \sigma(x_{n-3}) & \sigma^2(x_{n-4}) & & \gamma\sigma^{n-1}(x_{n-1}) \\
				x_{n-1} & \sigma(x_{n-2}) & \sigma^2(x_{n-3}) & \cdots & \sigma^{n-1}(x_0) \end{bmatrix}. 
\end{align*}

We refer to this representation as the \emph{left regular representation} of $\mc{C}$. The determinant and trace of $\rho(x)$ are referred to as the \emph{reduced norm} $\rnm(x)$ and \emph{reduced trace} $\rtr(x)$ of $x$, respectively. Note that we have the relations $\nm{\mc{C}/K}{x} = \rnm(x)^n$ and $\tr{\mc{C}/K}{x} = n\rtr(x)$. Here, $\nm{\mc{C}/K}{}$ and $\tr{\mc{C}/K}{}$ are similarly defined as the reduced norm and trace, but with respect to a basis of $\mc{C}$ over $K$.  

Given a cyclic algebra $\mc{C} = (L/K,\sigma,\gamma)$, we can determine whether $\mc{C}$ is division by means of $\gamma$. We recall here two results, which are used in Publications I, II and III.
\begin{lemma}
	Let $\mc{C} = (L/K,\sigma,\gamma)$ be a cyclic algebra of degree $n$. 
	\begin{itemize}
		\item[i)] \cite[Prop.~2.4.5]{hollanti:thesis}\quad If $\gamma$ is such that $\gamma^{\frac{n}{p}}\notin \Nm{L/K}{L^\times}$ for all primes $p \mid n$, then $\mc{C}$ is a division algebra. 
		\item[ii)] \cite[Thm.~7.1]{unger:quaternion}\quad Let $\mf{p}$ be a prime ideal of $\mc{O}_K$ with corresponding $\mf{p}$-adic valuation $\nu_{\mf{p}}$, and let $a \in K$ be such that $\nu_{\mf{p}}(a) = 1$. For any element $\gamma \in \mc{O}_K$ which is not a square $\bmod\ \mf{p}$, $\mc{C} = (K(\sqrt{a})/K,\sigma,\gamma)$ is a division algebra. 
	\end{itemize}
\end{lemma}

The element $\gamma$ is referred to as a \emph{non-norm element} for obvious reasons, and one should in fact think of $\gamma$ as an element in the quotient $K^\times/\nm{L/K}{L^\times}$. Note that $\gamma' = \nm{L/K}{\alpha}\gamma$ for some $\alpha \in L^\times$ if and only if $(L/K,\sigma,\gamma) \cong (L/K,\sigma,\gamma')$.

\subsection{Orders and Discriminants}
\label{subsec:orders}

Given a number field $K$, the collection of integral elements form the ring of integers $\mc{O}_K$ of $K$. This ring is the unique \emph{maximal order} of $K$, a concept which we will now recall in a more general context. 

Given a cyclic division algebra $\mc{C} = (L/K,\sigma,\gamma)$, an $\mc{O}_K$-\emph{order} $\Gamma$ in $\mc{C}$ is a subring of $\mc{C}$ sharing the same identity as $\mc{C}$ and such that $\Gamma$ is a finitely generated $\mc{O}_K$-module which generates $\mc{C}$ as a linear space over $K$. Maximality is defined with respect to inclusion, and every order is contained in a maximal order. 

Within a number field $K$, the ring of integers $\mc{O}_K$ is integrally closed and the unique maximal order of $K$. In general, a maximal order $\Gamma$ is not integrally closed, and a division algebra $\mc{C}$ may contain multiple maximal orders. In contrast, the following special order is often of interest due to its simple structure. It is in fact the initial source for space--time codes with non-vanishing determinants, and the main object of interest in Publication I. 
\begin{definition}
	Let $\mc{C} = (L/K,\sigma,\gamma)$ be a cyclic division algebra. The \emph{natural order} of $\mc{C}$ is the $\mc{O}_K$-module
	\begin{align*}
		\Gamma_{\mathrm{nat}} := \bigoplus\limits_{i=0}^{n-1}{e^i \mc{O}_L}.
	\end{align*}
\end{definition}
Note that $\Gamma_{\mathrm{nat}}$ is not closed under multiplication unless $\gamma \in \mc{O}_K$. 

Given an order $\Gamma$ in a cyclic division algebra $\mc{C}$, for every $g \in \Gamma\backslash\left\{0\right\}$ the reduced norm and trace satisfy $\rnm(g),\rtr(g) \in \mc{O}_K\backslash\left\{0\right\}$. Similarly to the number field case, we can define the \emph{discriminant} (ideal) of the order $\Gamma$ as the ideal 
\begin{align*}
	\disc(\Gamma/\mc{O}_K) := \langle \det\left(\rtr(x_i x_j)\right)^{n^2}\rangle
\end{align*} 
where $\left\{x_1,\ldots,x_{n^2}\right\} \in \Gamma^{n^2}$. When $\Gamma$ is a free $\mc{O}_K$-module, we can choose any basis of $\Gamma$ and view the discriminant as a number, rather than an ideal. All maximal orders of $\mc{C}$ share the same discriminant $\disc(\mc{C})$, referred to as the \emph{discriminant of the algebra}. Given two $\mc{O}_K$-orders $\Gamma_1 \subseteq \Gamma_2$, we have $\disc(\Gamma_2/\mc{O}_K) \mid \disc(\Gamma_1/\mc{O}_K)$, and consequently $\disc(\mc{C}) \mid \disc(\Gamma/\mc{O}_K)$ in $\mc{O}_K$ for any $\mc{O}_K$-order $\Gamma$. For the natural order $\Gamma_{\mathrm{nat}}$ we have $\disc(\Gamma_{\mathrm{nat}}/\mc{O}_K) = \disc(L/K)^n$. Hence, for any $F \subseteq K$ we have the important relation 
\begin{align*}
	\disc(\Gamma_{\mathrm{nat}}/\mc{O}_F) = \disc(L/F)^n\nm{K/F}{\gamma}^{n(n-1)}.
\end{align*}

\section{Lattices}
\label{sec:lattices}

The most important objects used in this thesis are lattices, structures which we introduce in this section alongside related notions and important properties. 

Consider a Euclidean space $E = (V,\langle \cdot,\cdot \rangle)$, consisting of a finite dimensional real vector space $V$ and an inner product. We have an isomorphism $E \cong \R^{\dm{E}}$, and we will henceforth consider the standard Euclidean space $\R^n$ with the standard inner product. 
\begin{definition}
	A \emph{lattice} $\Lambda \subset \R^n$ is the $\Z$-span of a set of vectors of $\R^n$, linearly independent over $\R$. 	
\end{definition}

Any lattice is isomorphic to $\Z^t$ as \emph{groups}, $t \le n$, and is thus a free abelian group of rank $\rk{\Lambda} = t$. We give an alternative group theoretic definition. 

\begin{definition}
	A \emph{lattice} $\Lambda \subset \R^n$ is a discrete subgroup of $\R^n$. 
\end{definition}

By discrete subgroup we mean that the metric on $\R^n$ defines the discrete topology on $\Lambda$. When $n = 1$, the situation is very simple, as any subgroup of $\R$ is either dense or discrete, while for $n > 1$, subgroups of $\R^n$ are not as easy to classify. A simple classification is however possible when only considering discrete subgroups. To be precise, for $\mb{b}_1,\ldots,\mb{b}_t$ $\R$-linearly independent vectors in $\R^n$, the subgroup $\mb{b}_1\Z +\cdots + \mb{b}_t\Z$ is discrete. Conversely, for any discrete subgroup $G$ of $\R^n$ there exist $\R$-linearly independent vectors $\mb{b}_1,\ldots,\mb{b}_t \in G$ such that $G = \mb{b}_1\Z + \cdots + \mb{b}_t\Z$.  

A lattice $\Lambda \subseteq \R^n$ can hence be expressed as a set 
\begin{align*}
	\Lambda = \left\{\left.\mb{x} = \sum\limits_{i=1}^{t}{\mb{b}_i z_i} \right| z_i \in \Z\right\},
\end{align*}
and we say that $\left\{\mb{b}_1,\ldots,\mb{b}_t\right\}$ forms a $\Z$-basis of $\Lambda$. It is often assumed that a lattice has full-rank in its ambient space, $\rk{\Lambda} = \dim(V)$. While this is not necessary for the definition, it can always be achieved by replacing the ambient space with the subset of $\R^n$ spanned by $\Lambda$. We will henceforth assume a lattice to be full unless stated otherwise, as the general case does not differ significantly. 

We can conveniently define a \emph{generator matrix} $M_{\Lambda} = \left(\mb{b}_i\right)_i$, so that every element of $\Lambda$ can be expressed as $\mb{x} = M_{\Lambda}\mb{z}$ for some $\mb{z} \in \Z^n$. The \emph{Gram matrix} of $\Lambda$ is $G_{\Lambda} = M_{\Lambda}^t M_{\Lambda}$.

To each lattice $\Lambda$ we can associate its \emph{fundamental parallelotope}, defined as $\mc{P}_{\Lambda} := \left\{\left. M_{\Lambda}\mb{y}\right| \mb{y} \in [0,1)^n \right\}$. Note that we can recover $\R^n$ as a disjoint union of the sets $\mb{x} + \mc{P}_{\Lambda}$ for all $\mb{x} \in \Lambda$. 
Since $M_{\Lambda}$ contains a $\Z$-basis of $\Lambda$, any change of basis is obtained via a unimodular matrix. Hence, the volume of the obtained parallelotope $\mc{P}_{\Lambda}$ is invariant under change of basis. Thus, we define the \emph{volume} of a lattice $\Lambda \subset \R^n$ as the Lebesgue measure of its fundamental parallelotope, 
\begin{align*}
	\vl{\Lambda} := \vl{\mc{P}_{\Lambda}} = \sqrt{\det(G_{\Lambda})}.
\end{align*} 

A subgroup $\Lambda' \subset \Lambda$ which is itself a lattice is called a \emph{sublattice} of $\Lambda$, and we refer to $\Lambda$ as a \emph{superlattice} for $\Lambda'$. If $\dm{\Lambda} = \dm{\Lambda'}$, the group index $|\Lambda/\Lambda'|$ is finite, and the volume of $\Lambda'$ is given by $\vl{\Lambda'} = \vl{\Lambda}|\Lambda/\Lambda'|$. 


A \emph{lattice quantizer} $Q_{\Lambda}: \R^n \to \Lambda$ is a function that maps each point in $\R^n$ to its closest lattice point $\mb{x} \in \Lambda$. For a lattice $\Lambda \subset \R^n$ and quantizer $Q_{\Lambda}$, the \emph{Voronoi cell} associated with an element $\mb{x} \in \Lambda$ is the set 
	\begin{align*}
		\mc{V}_{\Lambda}(\mb{x}) := \left\{\left. \mb{y} \in \R^n \right| Q_{\Lambda}(\mb{y}) = \mb{x} \right\}, 
	\end{align*}
and the \emph{basic Voronoi cell} of $\Lambda$ is $\mc{V}(\Lambda) = \mc{V}_{\Lambda}(\mb{0})$. 

While the volume of a lattice is independent of the basis, it is often desirable to have a basis consisting of short vectors. The squared norm of the shortest independent vectors of a lattice are known as the \emph{successive minima} $\lambda_1,\ldots,\lambda_n$ of $\Lambda$, which can be defined as 
\begin{align*}
	\lambda_i(\Lambda) := \left(\inf\left\{\left. r > 0 \right| \dim(B_{r} \cap \Lambda) \geq i \right\}\right)^2,
\end{align*} 
where $B_r$ denotes an $n$-sphere of radius $r$. Hence, $B_{\sqrt{\lambda_i}}$ is the smallest sphere containing $i$ linearly independent lattice points. 

Of special interest are the vectors $\mb{x} \in \Lambda$ satisfying $\|\mb{x}\|^2 = \lambda_1$, that is, the \emph{shortest vectors} in the lattice. The number of shortest vectors is also known as the \emph{kissing number} $\kappa(\Lambda)$ of $\Lambda$. In general, finding a shortest vector in an arbitrary lattice is only known to be NP-hard. The main classical result is due to Minkowski, and states that for a lattice $\Lambda$ of rank $n$, the first minimum satisfies $\lambda_1 \le n \vl{\Lambda}^{\frac{n}{2}}$. In particular, for every dimension there is an upper bound on the largest possible $\lambda_1$ over all lattices of equal volume, known as the \emph{Hermite constant}. 

As a particularly nice family of lattices, \emph{well-rounded} lattices contain a basis of the ambient space consisting exclusively of shortest vectors. In other words, a lattice in $\R^n$ is called \emph{well-rounded} if $\lambda_1 = \cdots = \lambda_n$.

\subsection{Theta Function and Flatness Factor}

%
%


Having introduced the basic concepts related to lattices, we now give the definition of a very important function. Given a lattice $\Lambda$, define for each $r \in \R$ the cardinality $\Omega_{\Lambda}(r) := \left|\left\{\left.\mb{x} \in \Lambda\ \right| \|\mb{x}\|^2 = r\right\}\right|$.
\begin{definition}
	Let $\Lambda \subset \R^n$ be a full lattice. The \emph{theta series} of $\Lambda$ is the generating function 
	\begin{align*}
		\Theta_{\Lambda}(q) = \sum\limits_{\mb{x} \in \Lambda}{q^{\|\mb{x}\|^2}} = 1+\sum\limits_{r>0}{\Omega_{\Lambda}(r)q^r}.
	\end{align*} 
\end{definition}

We remark that the theta series converges absolutely for $0 \le q < 1$, and is more generally defined in terms of a complex variable $q = e^{\pi i z}$, $z \in \C$. For the purposes of this thesis, however, it suffices to view $\Theta_{\Lambda}(q)$ as a formal power series in a real variable $q$. By giving the definition in terms of $\Omega_{\Lambda}(r)$, it is apparent that $\Theta_{\Lambda}(q)$ encodes important properties of $\Lambda$, \emph{e.g.}, 
\begin{align*}
	\argmin\limits_{r>0}\left\{\Omega_{\Lambda}(r) > 0 \right\} = \lambda_1; \quad \min\limits_{r>0}\left\{\Omega_{\Lambda}(r) > 0\right\} = \kappa(\Lambda).
\end{align*}

Although of great importance, the theta series is unfortunately only known in closed form for a handful of lattices, and is usually given in terms of the \emph{Jacobi theta functions} 
\begin{align}
	\theta_2(q) = \sum\limits_{i=-\infty}^{\infty}{q^{\left(i+\frac{1}{2}\right)^2}}, \qquad
	\theta_3(q) = \sum\limits_{i=-\infty}^{\infty}{q^{i^2}}, \qquad
	\theta_4(q) = \sum\limits_{i=-\infty}^{\infty}{(-q)^{i^2}}.
\end{align}

\begin{table}
\begin{center}
	\scalebox{0.86}{
	\begin{tabular}{|c||c|c|c|l|}
		\hline
		Lattice & Dimension & $\lambda_1$ & $\vl{\Lambda}$ & $\Theta_{\Lambda}(q)$ \\
		\hline \hline
		$\Z^n$ & $n \ge 1$ & 1 & 1 & $\theta_3^n(q)$ \\
		\hline
		$D_n$ & $n \ge 3$ & 2 & 2 & $\frac{1}{2}(\theta_3^n(q)+\theta_4^n(q))$ \\
		\hline
		$A_2$ & 2 & 1 & $\sqrt{3/4}$ & $\theta_2(q)\theta_2(q^3)+\theta_3(q)\theta_3(q^3)$ \\
		\hline
		$E_8$ & 8 & 2 & 1 & $\frac{1}{2}(\theta_2^8(q)+\theta_3^8(q)+\theta_4^8(q))$ \\
		\hline
		$K_{12}$ & 12 & 4 & 27 & \vtop{\hbox{\strut $\frac{9}{32}\theta_2^6(q)\theta_2^6(q^3)+\left(\theta_2(q^4)\theta_2(q^{12})+\theta_3(q^4)\theta_3(q^{12})\right)^6$}\hbox{\strut $+ \frac{45}{16}\theta_2^4(q)\theta_2^4(q^3)\left(\theta_2(q^4)\theta_2(q^{12})+\theta_3(q^4)\theta_3(q^{12})\right)^2$}} \\
		\hline
		$\Lambda_{24}$ & 24 & 4 & 1 & $\frac{1}{2}(\theta_2^8(q)+\theta_3^8(q)+\theta_4^8(q))^3-\frac{45}{16}(\theta_2(q)\theta_3(q)\theta_4(q))^8$ \\
		\hline 
	\end{tabular}}
	\caption{Various important lattices and their basic attributes.}	
	\label{tab:lattices}
\end{center}
\end{table}

Even so, the Jacobi theta functions are rather complicated. The reason for this small set of lattices with known closed-form theta series is that, similarly to finding short vectors, efficient computing of lattice points with a given length in arbitrary domains and arbitrary dimensions is still an open problem. While many results have been obtained over the last two decades, for example the remarkable work by Fukshansky and Schürmann \cite{fukshansky:bounds_frobenius} or Widmer \cite{widmer:counting_lattice_points}, the results are of such a general nature that the upper bounds are far from being tight, even for very structured lattices and domains. Thus, being able to efficiently compute even an approximation of the theta series of an arbitrary lattice is a problem which is interesting in its own right. The following result is derived in Publication IV, additionally motivated by practical applications. 
\begin{theorem}
\label{thm:theta_approx}
Let $\Lambda \subset \R^n$ be a full lattice with volume $\vl{\Lambda}$ and minimal norm $\lambda_1$. For $0 \le q < 1$, the theta series $\Theta_{\Lambda}(q)$ can be expressed as 
	\begin{align*}
		\Theta_{\Lambda}(q) = (1-q^{\lambda_1}) - \frac{\log(q)\lambda_1^{\frac{n}{2}+1}\pi^{\frac{n}{2}}}{\Gamma\left(\frac{n}{2}+1\right)\vl{\Lambda}}\int\limits_{1}^{\infty}{t^{\frac{n}{2}}q^{\lambda_1 t} dt} + \Xi(\Lambda,n,L,q),
	\end{align*}
with an error term $\Xi(\Lambda,n,L,q)$ involving a constant which depends on $n$, $\Lambda$, and a Lipschitz constant $L$.
\end{theorem}

Finally, we can introduce the notion of the \emph{flatness factor} of a lattice, which is a direct function of its theta series, and relevant in Publications IV and VII. Consider the usual $n$-dimensional zero-mean Gaussian PDF with variance $\sigma^2$, 
\begin{align*}
	f(\mb{t},\sigma^2) = \frac{1}{(\sqrt{2\pi\sigma^2})^n} e^{-\frac{\|\mb{t}\|^2}{2\sigma^2}}.
\end{align*}
We let the variable $\mb{t}$ range over points over a (possibly shifted) full lattice $\Lambda$, yielding for $\mb{y} \in \R^n$ the sum of Gaussian functions 
\begin{align*}
	f(\Lambda+\mb{y}, \sigma^2) := \sum\limits_{\mb{x} \in \Lambda}{f(\mb{x}+\mb{y},\sigma^2)}. 
\end{align*}
As a function of $\mb{y}$, $f(\Lambda+\mb{y})$ is $\Lambda$-periodic, and defines a PDF on the basic Voronoi cell $\mc{V}(\Lambda)$, which we refer to as the \emph{lattice Gaussian PDF}. For the centered function $f(\Lambda,\sigma^2)$, we have the useful identity 
\begin{align*}
	f(\Lambda,\sigma^2) &= \sum\limits_{\mb{x} \in \Lambda}{f(\mb{x},\sigma^2)} = \frac{1}{(\sqrt{2\pi\sigma^2})^n}\sum\limits_{\mb{x} \in \Lambda}{e^{-\frac{\|\mb{x}\|^2}{2\sigma^2}}} \\ 
	&= \frac{1}{(\sqrt{2\pi\sigma^2})^n}\Theta_{\Lambda}\left(e^{-\frac{1}{2\sigma^2}}\right).
\end{align*}

\begin{definition}
	Let $\Lambda \subset \R^n$ be a full lattice, and for $\mb{y} \in \R^n$, let $f(\Lambda+\mb{y},\sigma^2)$ denote the lattice Gaussian PDF of the lattice $\Lambda+\mb{y}$. The \emph{flatness factor} of $\Lambda$ is defined as 
	\begin{align*}
		\varepsilon_{\Lambda}(\sigma^2) := \max\limits_{\mb{y} \in \R^n}\left|\frac{f(\Lambda+\mb{y},\sigma^2)}{1/\vl{\Lambda}}-1\right|.
	\end{align*}	
\end{definition}

The \emph{flatness factor} is a quantity which measures the deviation of the lattice Gaussian PDF from the uniform distribution on the Voronoi cell $\mc{V}(\Lambda)$. It is easy to show that the maximum is achieved for $\mb{y} \in \Lambda$, and as an immediate consequence we can relate the flatness factor to the theta series of $\Lambda$ by the equation
\begin{align*}
 	\varepsilon_{\Lambda}(\sigma^2) = \frac{\vl{\Lambda}}{(\sqrt{2\pi\sigma^2})^n}\Theta_{\Lambda}\left(e^{-\frac{1}{2\sigma^2}}\right)-1.
\end{align*}

\subsection{General Lattices}
\label{subsec:gen_lattices}

We have defined lattices as discrete subgroups of $\R^n$ and they are, by definition, free $\Z$-modules. It is however possible and often desirable to extend the definition to other rings and ambient spaces, such as the ring of integers of a number field, or an order in a cyclic division algebra. In this more general context, we define a lattice $\Lambda$ to be a discrete and finitely generated abelian subgroup of a real or complex ambient space $V$. In the previous derivations, we have set $V = \R^n$. When $V = \C^n$ or $V = \mat(n,\C)$, we first need to replace the standard inner product with the Hermitian form $\langle \mb{x},\mb{y} \rangle = \mb{x}^\dagger \mb{y}$ in the former, and by $\langle X,Y \rangle = \tr{}{X^\dagger Y}$ in the latter case, where $\dagger$ denotes conjugate transpose. In these cases, we can also identify a lattice in $V$ with a lattice in $\R^{2n}$ or $\R^{2n^2}$, respectively, via the $\R$-linear isometry 
\begin{equation}
\label{eqn:isometry}
\resizebox{.9\textwidth}{!}{$
	\iota: 
	\begin{cases} 
		\C^n \to \R^{2n}; & (u_1,\ldots,u_n)^t \mapsto \left(\Re(u_1),\Im(u_1),\ldots,\Re(u_n),\Im(u_n)\right)^t, \\ 
		\mat(n,\C) \to \R^{2n^2}; & (\mb{u}_1,\ldots,\mb{u}_n) \mapsto \left(\Re(u_{11}),\Im(u_{11}),\ldots,\Im(u_{1n}),\ldots,\Re(u_{nn}),\Im(u_{nn})\right)^t.
	\end{cases}$}
\end{equation}

We have $\|\mb{u}\| = \|\iota(\mb{u})\|$ and $\|U\|_F = \|\iota(U)\|$, respectively, where $\|\cdot\|_F$ denotes the Frobenius norm, and $\iota$ maps full lattices in $V$ to full lattices in the target Euclidean space. 

For $V = \C^n$ it is straightforward to define most of the notions. For $V = \mat(n,\C)$, let $\Lambda \subset V$ be a full lattice with $\Z$-basis $\left\{B_1,\ldots,B_n\right\}$, $B_i \in V$. A generator matrix and the corresponding Gram matrix for $\Lambda$ can be given as 
\begin{align*}
	M_{\Lambda} = \left(\iota(B_i)\right)_i; \quad G_{\Lambda} = M_{\Lambda}^{\dagger} M_{\Lambda} = \left(\Re(\tra(B_i^\dagger B_j))\right)_{i,j}.
\end{align*}
The volume of $\Lambda$ is the volume of the corresponding lattice $\iota(\Lambda)$ in $\R^{2n^2}$, \emph{i.e.}, $\vl{\Lambda} = \sqrt{\det(G_{\Lambda})}$.

\chapter{Lattice Codes and the Physical Layer}
\label{chp:plc}

In the previous chapter we introduced the most important mathematical tools used in this thesis. We now give a brief overview on the physical medium and related code design. The main focus in this chapter lies on the basic wireless point-to-point channel model and related statistical quantities. We do not further specify transmission protocols, as the fundamental principles of wireless transmission models and characteristics of wireless channels can be illustrated with this most basic setup. More specialized settings will be discussed later in Chapter~\ref{chp:communications} when we review the results of the publications constituting this thesis.

Further, we review the construction and properties of two types of lattice codes, and recall the related code design criteria. The main references for this chapter are \cite{hollanti:thesis}, \cite{jafarkhani:stc} and \cite{zamir:lattices}.

\section{Fading Channel Model}
\label{sec:channel}

In a wireless environment, in contrast to wired channels, a transmitted signal is not bound to follow a specific path from the transmitter to the receiver. Indeed, many different paths exist, and consequently different electromagnetic waves will coalesce at the receiver, causing a superimposed channel output. Together with dissipation effects caused by, \emph{e.g.}, urban structures, as well as interference, the signal experiences \emph{fading}. Various statistical models exist to describe this phenomenon, two important ones being the \emph{Rayleigh} and \emph{Rician} fading model. The latter is especially useful when a (partial) line of sight is assumed between the transmitter and receiver, \emph{i.e.}, when there is a clearly dominant signal. This is however not necessarily a realistic assumption in general urban environments or long-distance communications, and we will consider the former fading model. 

In addition to fading, thermal \emph{noise} at the receiver further distorts the channel output. This additional noise term is usually modeled as a zero-mean Gaussian random variable with finite variance.

Let us describe a Rayleigh fading channel model with additive white Gaussian noise. Assume a single source equipped with $n_t \ge 1$ transmit antennas and a single destination with $n_r \ge 1$ receive antennas. To combat the destructive effects of fading, the transmitter encodes its data into a codeword (matrix) $X \in \mat(n_t\times T, \C)$, where $T$ denotes the number of channel uses, also referred to as \emph{delay}. That is, the source communicates to the destination over $T$ subsequent time slots. Here, each column of $X$ corresponds to the signal vector transmitted in each time slot, across the available transmit antennas. This strategy of providing multiple independently fading copies of the same signal to the receiver is known as enabling \emph{diversity}. More concretely, we refer to spatial and temporal diversity when using multiple antennas and time slots, respectively; we code over \emph{space and/or time}. 

The physical channel is modeled by a random matrix $H = \left(h_{ij}\right)_{i,j} \in \mat(n_r \times n_t,\C)$, which is assumed to remain static for $T$ time slots and then change independently of its previous state. The entries of $H$ are complex variables with i.i.d. normal distributed real and imaginary parts, 
\begin{align*}
	\Re(h_{ij}), \Im(h_{ij}) \sim \mc{N}(0,\sigma_h^2),
\end{align*} 
yielding a Rayleigh distributed envelope 
\begin{align*}	
	|h_{ij}| = \sqrt{\Re(h_{ij})^2 + \Im(h_{ij})^2} \sim \mathrm{Ray}(\sigma_h) 
\end{align*}
with scale parameter $\sigma_h$, which gives this fading model its name. 

The additive noise is modeled by a matrix $N \in \mat(n_r\times T,\C)$ with i.i.d. complex Gaussian entries with zero-mean and variance $\sigma_n^2$. In summary, the channel output at the receiver can be modeled by the well-known channel equation 
\begin{align*}
	Y = HX + N.
\end{align*}

For terminology, we differentiate the cases where $n_t, n_r \ge 2$, to which we refer to as the \emph{multiple-input multiple-output} (MIMO) model, as well as the special case $(n_t,n_r) = (1,1)$, the \emph{single-input single-output} (SISO) channel model. The receiver is usually assumed to have channel state information, that is, to know the channel matrix $H$, which is especially important for decoding purposes. In this thesis, the transmitter is assumed to know the statistics of $H$, but not know the current channel realization.

As the fading and noise matrices need to be treated as random variables, the resulting performance depends entirely on the employed code, \emph{i.e.}, the finite set from which the codewords $X$ are taken. We present two different types of codes whose construction and performance rely on an underlying lattice structure.

\section{Lattice-Based Coding for Wireless Channels}
\label{sec:coding}

After discussing the basic properties of a typical channel model in the previous section, we now move on to study the construction of lattice codes from algebraic structures. We start with an introduction to \emph{space--time coding}, a technique employed in MIMO communications, which we consider in Publications I, II and III, and recall the related code design criteria. We then review the construction and properties of \emph{nested lattice codes} used, \emph{e.g.}, in the communication setups considered in Publications IV, V, VI and VII.

\subsection{Space--Time Codes}
\label{subsec:stc}
 
Recall the Rayleigh fading MIMO channel model. For communication over $n_t$ transmit antennas and $T$ time instances, we have seen that codewords $X$ need to be taken from some finite collection of matrices $\mc{X} \subset \mat(n_t\times T,\C)$. Naively, we could define a code to simply be a finite collection of such matrices. However, the so-called \emph{coding gain}, a quantity related to the minimum determinant of the code, vanishes for non-discrete structures. We will define these notions shortly. To circumvent this problem, forcing a discrete structure on the code is helpful, \emph{e.g.}, a lattice structure.
\begin{definition}
\label{def:stc}
	Let $\left\{ B_i \right\}_{i=1}^{k}$ be an independent set of fixed $n_t \times T$ complex matrices. A \emph{linear space--time block code} of rank $k$ is a set of the form
\begin{align*}
	\mc{X} = \left\{\left. \sum\limits_{i=1}^{k}{B_i s_i} \right| s_i \in S \right\},
\end{align*}
where $S \subset \Z$ is the finite \emph{signaling alphabet} used. 

If the matrices $\left\{ B_i \right\}_{i=1}^{k}$ form a basis of a \emph{lattice} $\Lambda \subset \mat(n_t\times T, \C)$, then $\mc{X}$ is called a \emph{space--time lattice code} of rank $k = \rk{\Lambda}$. 
\end{definition}

We henceforth refer to such a code $\mc{X}$ simply as a space--time code. We identify the lattice $\Lambda$ underling the code $\mc{X}$ with the corresponding lattice $\iota(\Lambda) \subset \R^{2n_t T}$ (cf. \eqref{eqn:isometry}), and \emph{carve out} the finite code $\mc{X}$ from the infinite lattice by fixing a bounding region $D \subset \R^{2n_t T}$ and setting $\iota(\mc{X}) = \iota(\Lambda) \cap D$. The \emph{code size} is the cardinality of the finite set of codewords. As the transmit power consumption is directly related to the Frobenius norm of the transmitted codeword, the bounding region $D$ should be chosen such that for a fixed code size, the elements of $\iota(\mc{X})$ are efficiently packed inside $D$, and maintain a low average power. Typical choices are a spherical shape, which yields the best packing at the cost of more complex bit labeling, and cubic shapes, which can be easily labeled.

The \emph{code rate} of $\mc{X}$ is defined as $R = k/T$ real symbols per channel use. For $n_r$ receive antennas, the code is said to be \emph{full-rate} if $R = 2 n_r$. Here, full-rate is defined as the maximum rate that still maintains the discrete structure at the receiver and allows for linear detection methods such as sphere-decoding. 

Consider a space--time code $\mc{X}$, and let $X \in \mc{X}$ be the transmitted codeword. \emph{Maximum likelihood} (ML) decoding refers to the maximization of the conditional probability of receiving $Y$ when $X$ was transmitted, assuming the channel realization $H$. As the receiver has channel state information and the noise is assumed to be zero-mean, ML decoding in this simple communication setup can equivalently be carried out by computing an estimate of the transmitted codeword as
\begin{align}
\label{eqn:ml_decod}
	\hat{X} = \argmin\limits_{X \in \mc{X}}{\|Y-HX\|_F^2}.
\end{align}
 
Let us assume $n_t \ge T$. The probability $\Pr(X \to X')$ that a codeword $X' \neq X$ is decoded when $X$ was sent is asymptotically upper bounded with increasing \emph{signal-to-noise ratio} ($\snr$) as 
\begin{align*}
	\Pr(X \to X') \le \left(\det\left((X-X')(X-X')^\dagger\right)\snr^{n_t}\right)^{-n_r}.
\end{align*}

From this upper bound, two design criteria can be derived \cite{tarokh:stc}. The \emph{diversity gain} of a code is the asymptotic slope of the error probability curve with respect to the $\snr$ in a $\log-\log$ scale, and relates to the minimum rank $\rk{X-X'}$ over all pairs of distinct code matrices $(X,X') \in \mc{X}^2$. Thus, the minimum rank of $\mc{X}$ should ideally satisfy  
\begin{align*}
	\min_{X \neq X'} \rk{X-X'} = \min\{n_t,T\}. 
\end{align*}
A code satisfying this criterion is called a \emph{full-diversity} code. 

On the other hand, the \emph{coding gain} measures the difference in $\snr$ required for two different codes to achieve the same error probability. For a full-diversity code, this is proportional to the determinant 
\begin{align*}
	\det\left((X-X')(X-X')^\dagger\right). 
\end{align*}
As a consequence, the minimum determinant over all pairs of codewords $(X,X') \in \mc{X}^2$,
\begin{align*}
		\min_{X \neq X'} \det\left((X-X')(X-X')^{\dagger}\right), 
\end{align*}
should be as large as possible. For the infinite code
\begin{align*}
	\mc{X}_\infty = \left\{\left.\sum\limits_{i=1}^{k}{s_i B_i} \right| s_i \in \Z \right\}
\end{align*}
we define the \emph{minimum determinant} as the infimum
\begin{align*}
 	\Delta_{\min}(\mc{X}_{\infty}) := \inf_{X \neq X'} \det\left((X-X')(X-X')^{\dagger}\right),
\end{align*}
and if $\Delta_{\min}(\mc{X}_\infty) > 0$, \emph{i.e.}, the determinants do not vanish as the code size increases, the code is said to have the \emph{non-vanishing determinant} property. 

Let hereinafter $n_t = T =: n$. Given a lattice $\Lambda \subset \mat(n,\C)$, we have by linearity
\begin{align*}
	\Delta_{\min}(\Lambda) = \inf\limits_{0\neq X \in \Lambda}{\left|\det(X)\right|^2},
\end{align*}
and we call $\Delta_{\min}(\Lambda)$ the \emph{minimum determinant} of the lattice. This implies that any lattice $\Lambda$ with the non-vanishing determinant property can be scaled so that $\Delta_{\min}(\Lambda)$ achieves any wanted positive value. Consequently, the comparison of two different lattices requires some sort of normalization. Let $\Lambda$ be a full lattice with volume $\vl{\Lambda}$ and Gram matrix $G_{\Lambda}$.
The \emph{normalized minimum determinant} and \emph{normalized density} of $\Lambda$ are the normalized quantities
\begin{align*}
	\delta(\Lambda) = \frac{\Delta_{\min}(\Lambda)}{\vl{\Lambda}^{\frac{1}{2n}}}; \quad 
	\eta(\Lambda) = \frac{\Delta_{\min}(\Lambda)^{2n}}{\vl{\Lambda}},
\end{align*}
and satisfy the relation $\delta(\Lambda)^2 = \eta(\Lambda)^{\frac{1}{n}}$. Thus, for fixed minimum determinant, the coding gain can be increased by maximizing the density of the code, or equivalently the density of the lattice, a problem which translates into a discriminant minimization problem \cite{vehkalahti:dense_mimo}, as we shall see shortly. 

We illustrate how to cleverly design space--time codes satisfying the two desired criteria. In \cite{sethuraman:stc} it was observed that for a field $K$ and division $K$-algebra $\mc{D}$, any finite subset $\mc{X}$ of the image of a ring homomorphism $\phi: \mc{D} \mapsto \mat(n, K)$ satisfies $\rk{X-X'} = n$ for any distinct $X, X' \in \mc{X}$. This leads to a straightforward approach for constructing full-diversity codes, namely by choosing the underlying structure to be a division algebra. In the same article, cyclic division algebras were proposed for code construction as a particular class of division algebras. The ring homomorphism $\phi$ is the link between the division algebra and a full-diversity space--time code. 

Let $\mc{C} = (L/K,\sigma,\gamma)$ be a cyclic division algebra of degree $n$. The left-regular representation $\rho: \mc{C} \to \mat(n,\C)$ is an injective ring homomorphism. We identify elements in $\mc{C}$ with elements in $\mat(n,\C)$ via $\rho$, and can define a space--time code to be a finite subset of $\rho(\mc{C})$ or $\rho(\mc{C})^t$. While this definition is sufficient to ensure full-diversity, a lattice structure can be additionally enforced by restricting the choice of elements to an order $\Gamma \subset \mc{C}$. Therefore, we carve a space--time code as a collection of short vectors $\mc{X} \subset \rho\left(\Gamma\right)$, or its transpose. To be consistent with Definition~\ref{def:stc}, let $\rk{\Lambda} = 2n^2$ and let $\left\{ B_i \right\}_{i=1}^{2n^2}$ be a matrix basis for $\Gamma$ over $\Q$. For a fixed signaling alphabet $S \subset \Z$, symmetric around the origin, the space--time code $\mc{X}$ is of the form
\begin{align*}
	\mc{X} = \left\{\left. \sum\limits_{i=1}^{2n^2}{s_i B_i} \right| s_i \in S \right\}. 
\end{align*}

Recall that for every $c \in \Gamma$ we have $\det(\rho(c)) \in \mc{O}_{K}$. When $K = \Q$ or $K$ is imaginary quadratic, this implies $\det(\rho(c)) \ge 1$, thus guaranteeing non-vanishing determinants. This was first achieved for codes based on cyclic algebras in \cite{belfiore:golden} for the \emph{Golden code} by restricting the matrix entries to the ring of integers of the center \cite{belfiore:nvd, rekaya:nvd}, and the results were generalized to other \emph{Perfect codes} in \cite{oggier:perfect, elia:perfect}. The usefulness of orders more generally, however, was first noticed in \cite{hollanti:order1}. 

We finally relate the minimum determinant of a code to the density of the lattice. If the center $K$ of the cyclic division algebra is quadratic imaginary and the order $\Gamma$ admits an $\mc{O}_K$-basis, then $\rho(\Gamma)$ is a lattice and the measure of the fundamental parallelotope $\mc{P}(\rho(\Gamma))$ is directly related to the discriminant of the order $\Gamma$ as \cite{vehkalahti:dense_mimo}
\begin{align*}
	\vl{\rho(\Gamma)} = c(K,n)|\disc(\Gamma/\mc{O}_K)|,
\end{align*}
where $c(K,n)$ is a constant which depends on the center and extension degree. Thus, in order to maximize the density of a lattice arising from an order in a cyclic division algebra, the discriminant of the order needs to be minimized. This observation is crucial and is the main motivation behind our work in Publication I.

\subsubsection{Fast-Decodable Codes}
\label{subsubsec:fd_codes}

We have just seen what properties a space--time code should exhibit to potentially ensure a good performance, at least in terms of reliability. There are however more aspects of the communication process which need to be taken into consideration. For example, a complicated lattice structure makes it more complex to encode a signal. In contrast, orthogonal lattices are worse for coding gain purposes, but allow for efficient bit-labeling and somewhat lower complexity encoding. On the receiver's side, the structure of the code lattice determines the complexity of the decoding process. Indeed, the major bottle-neck in effective implementation of algebraic space--time codes has traditionally been their decoding complexity. The concept of fast decodability was introduced in \cite{biglieri:fd} in order to address the possibility for reducing the dimension of the ML decoding problem (cf. \eqref{eqn:ml_decod}) without having to resort to suboptimal decoding methods.

The ML decoding complexity of a rank-$k$ space--time code $\mc{X}$ is defined as the minimum number of values that have to be computed for finding the solution to \eqref{eqn:ml_decod}. The upper bound is the worst-case complexity $|S|^k$ corresponding to an exhaustive search, where $S \subset \Z$ is the finite signaling alphabet. The following definition is hence straightforward. 	
\begin{definition}
	A space--time code $\mc{X}$ is said to be \emph{fast-decodable} if its worst-case ML decoding complexity is $|S|^{k'}$ for $k' < k-2$. 
\end{definition}
This complexity is given in real dimensions, of which two can be eliminated by simple Gram-Schmidt orthogonalization. Hence, we require a strict inequality. The term \emph{fast-decodable} is however somewhat misleading. While a fast-decodable space--time code exhibits a reduced worst-case decoding complexity, even a significant reduction can result in a decoding process which is too slow for practical use.

We proceed to investigate how to determine the decoding complexity of a space--time code $\mc{X}$. Let $\left\{B_i\right\}_{i=1}^k$ be a basis of $\mc{X}$, and $X \in \mc{X}$ the transmitted signal. For $H$ the channel matrix and $\iota$ the isometry from \eqref{eqn:isometry}, define the matrix $B = \left(\iota(HB_i)\right)_i \in \mat(2T n_r\times k, \R)$. The equivalent received codeword can be expressed as $\iota(HX) = B\mb{s}$ for a coefficient vector $\mb{s}^t = (s_1,\ldots,s_k) \in S^k$. Performing $QR$-decomposition on $B$, $B = QR$ with $Q$ unitary and $R$ upper triangular, we get an equivalent decoding problem which requires to solve 
\begin{align*}
	\hat{\mb{s}} = \argmin\limits_{\mb{s} \in S^k}{\|\iota(Y) - B\mb{s}\|^2} = \argmin\limits_{\mb{s} \in S^k}{\|Q^\dagger \iota(Y) - R\mb{s}\|^2},
\end{align*}
a problem which can be solved using a real sphere decoder. It is now clear that the structure of the matrix $R$ influences the complexity of decoding. With zero entries at specific places, the involved variables can be decoded independently of each other, allowing for parallelization and reducing the decoding complexity. We make this more explicit by refining the definition of fast-decodability. 

\begin{definition}
	A space--time code $\mc{X}$ is called
	\begin{itemize}
		\item[i)] \emph{conditionally $g$-group decodable} if there exists a partition of $\left\{ 1,\ldots,k \right\}$ into $g+1$ non-empty subsets $\left\{\Gamma_1,\ldots,\Gamma_g,\Gamma^{\mc{X}}\right\}$ with $g \ge 2$, such that $B_i B_j^{\dagger} + B_j B_i^{\dagger} = 0$ for $i \in \Gamma_u$, $j \in \Gamma_v$ and $1 \le u < v \le g$. 
	
		\item[ii)] \emph{$g$-group decodable} if there exists a partition of $\left\{ 1,\ldots,k\right\}$ into $g$ non-empty subsets $\left\{\Gamma_1,\ldots,\Gamma_g\right\}$ such that $B_i B_j^{\dagger} + B_j B_i^{\dagger} = 0$ for $i \in \Gamma_u$, $j \in \Gamma_v$, and $u \neq v$. 
	\end{itemize}
\end{definition}

\begin{remark}
	These refined definitions are sufficient for the work carried out in Publications II and III, though it is noteworthy that the definition of these types of fast-decodable codes is usually based on conditions derived from a so-called \emph{Hurwitz-Radon quadratic form} approach. In recent work \cite{mejri:fd_revisited}, Mejri \emph{et al.} showed that this typical approach does not capture all families of codes with reduced ML-decoding complexity. 
\end{remark}

The advantage of this refined definition is that, after possibly reindexing the basis matrices, the $R$-matrix obtained for conditionally $g$-group decodable and $g$-group decodable codes has the particular form
\begin{align*}
	R = \begin{bmatrix} D_1 & & & N_1 \\ & \ddots & & \vdots \\ & & D_g & N_g \\ & & & N \end{bmatrix}\ \text{ or }\ R = \begin{bmatrix} D_1 & & \\ & \ddots & \\ & & D_g \end{bmatrix},
\end{align*}
respectively, where the blocks $D_i$ are of size $|\Gamma_i|\times|\Gamma_i|$ and $N$ is a square upper-triangular $\left|\Gamma^{\mc{X}}\right|\times\left|\Gamma^{\mc{X}}\right|$ matrix \cite{berhuy:fd}. Here, the empty slots denote zero entries.

Conditionally $g$-group decodable and $g$-group decodable codes are examples of families of fast-decodable space--time codes. The refined definitions, however, allow one to deduce the exact decoding complexity reduction. Decoding the last $\left|\Gamma^{\mc{X}}\right| \ge 0$ variables gives a complexity of $|S|^{\left|\Gamma^{\mc{X}}\right|}$, while the remaining variables can be decoded in $g$ parallel steps, with step $i$ involving $|\Gamma_i|$ variables. Thus, the explicit worst-case ML decoding complexity of a (conditionally) $g$-group decodable space--time code $\mc{X}$ is \cite{jithamithra:sphere_decod}
\begin{align*}
	|S|^{\left|\Gamma^{\mc{X}}\right| + \max\limits_{1 \le i \le g}{|\Gamma_i|}}.
\end{align*}

\subsection{Nested Lattice Codes}
\label{subsec:nested_codes}

Significantly older than the concept of a space--time code is that of a \emph{Voronoi code}, introduced in \cite{conway:voronoi}, and hereinafter referred to as a \emph{nested lattice code}.
Given a pair of $n$-dimensional full lattices $\Lambda_C \subseteq \Lambda_F \subset \R^n$, we say that $\Lambda_C$ is \emph{nested} in $\Lambda_F$, and refer to $\Lambda_F$ as the \emph{fine} lattice, and to $\Lambda_C$ as the \emph{coarse} lattice. The generator matrices of two nested lattices are related in a simple manner, namely $M_{\Lambda_C} = M_{\Lambda_F}G$ for $G \in \mat(n,\Z)$ of determinant $\det(G) \ge 1$. Similarly, a sequence $\Lambda_1,\ldots,\Lambda_s$ of lattices is \emph{nested} if $\Lambda_1 \subseteq \Lambda_2 \subseteq \cdots \subseteq \Lambda_s$. The idea behind nested lattice codes is to construct a finite codebook as the set of representatives of the quotient group $\Lambda_F/\Lambda_C$. Let us make this more explicit.  
\begin{definition}
	Let $\Lambda_C \varsubsetneq \Lambda_F$ be a pair of properly nested lattices. A \emph{nested lattice code} $\mc{C}(\Lambda_C,\Lambda_F)$ is the set of representatives
	\begin{align*}
		\mc{C}(\Lambda_C,\Lambda_F) := \left\{\left.\left[\mb{x}\right] \in \Lambda_F\ (\bmod\ \Lambda_C) \right| \mb{x} \in \Lambda_F \right\} = \Lambda_F \cap \mc{V}(\Lambda_C).	
	\end{align*}
\end{definition}

\begin{figure}[ht]
	\includegraphics[trim={11 2cm 17cm 6cm},clip,width=.65\textwidth]{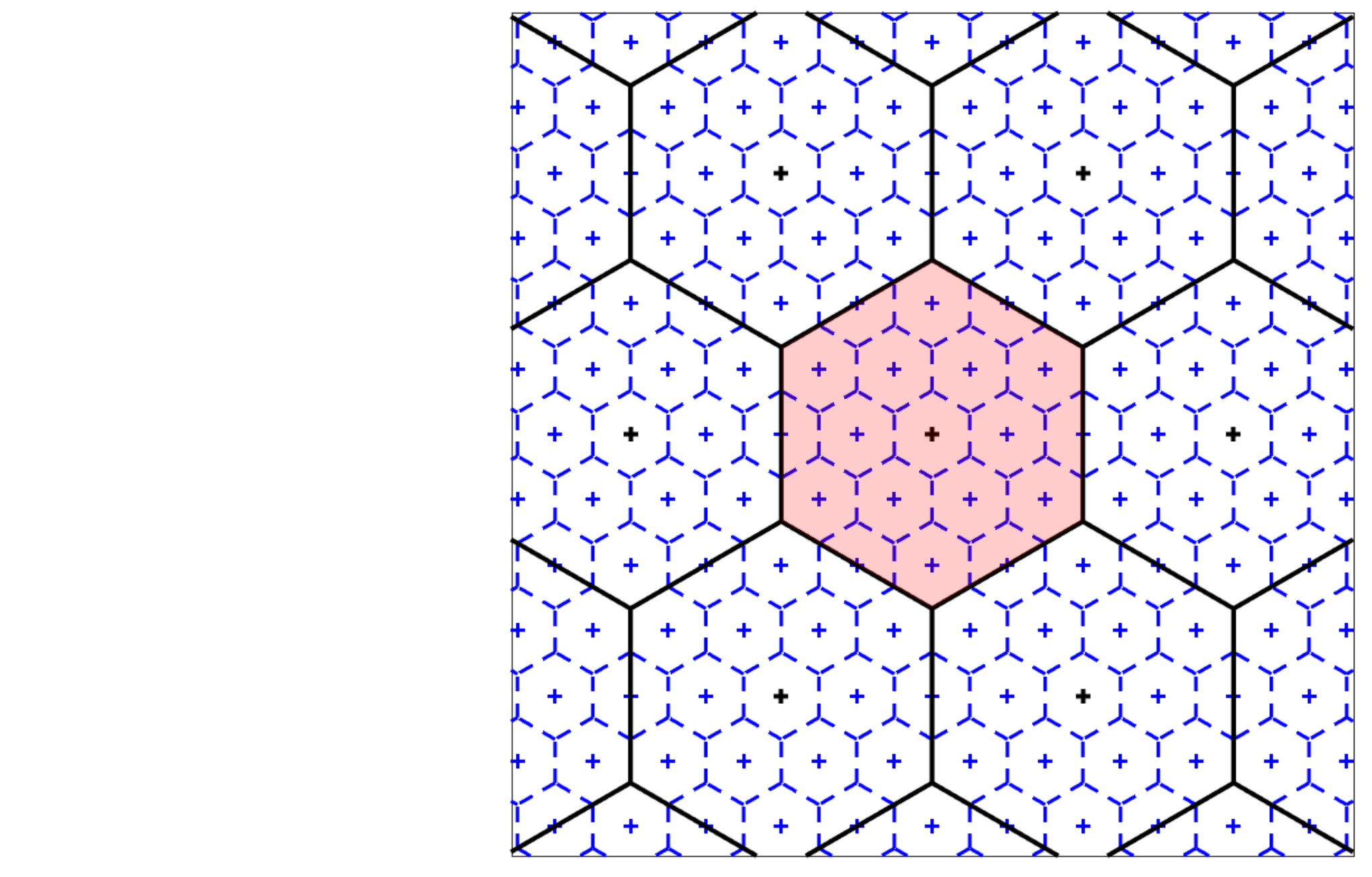}
	\caption{Nested lattices $A_2 = \Lambda_C \subset \Lambda_F = 4\Lambda_C$ with the Voronoi cells around each lattice point of the coarse (solid) and fine (dashed) lattices. \newline
The centered Voronoi cell $\mc{V}(\Lambda_C)$ (red) contains a set of representatives for a nested lattice code $\mc{C}(\Lambda_C,\Lambda_F)$ of cardinality $|\mc{C}(\Lambda_C,\Lambda_F)| = \left|\Lambda_F/\Lambda_C\right| = 16$.}
	\label{fig:nested_code}
\end{figure}

We illustrate the concept of a nested lattice code in Figure~\ref{fig:nested_code}. 
The elements in $\Lambda_F \cap \mc{V}(\Lambda_C)$ are called the \emph{coset leaders} of $\Lambda_C$, of which there are $\left|\Lambda_F \cap \mc{V}(\Lambda_C)\right| = \left[\Lambda_F:\Lambda_C\right] = \det(G)$. In Figure~\ref{fig:nested_code}, it is visible that some elements of the fine lattice lie on the border of $\mc{V}_C$. Those are mapped to coset leaders in a systematic fashion and in such a way that the shifted \emph{cosets} $[\mb{x}] + \Lambda_C$ are disjoint, where $[\mb{x}] \in \Lambda_F \cap \mc{V}(\Lambda_C)$.

Given a nested lattice code $\mc{C}(\Lambda_C,\Lambda_F)$, the \emph{code rate} (in bits per dimension) is defined as 
\begin{align*}
	\mc{R} = \frac{1}{n}\log{\left(\left|\mc{C}(\Lambda_C,\Lambda_F\right|\right)} = \frac{1}{n}\log{\left(\frac{\vl{\Lambda_C}}{\vl{\Lambda_F}}\right)} = \frac {1}{n}\log\left(\left|\Lambda_F/\Lambda_C\right|\right).
\end{align*}

We move on to give an overview of the \emph{goodness} of nested lattices for coding purposes. Consider the \emph{normalized second moment} of a lattice $\Lambda \subset \R^n$, defined as
\begin{align*}
	\sigma^2_{\Lambda} := \frac{1}{n\vl{\Lambda}^{1+\frac{2}{n}}}\int_{\mc{V}(\Lambda)}{\|\mb{t}\|^2 d\mb{t}}.
\end{align*}
The normalized second moment $\sigma^2_{B^{(n)}}$ of an $n$-sphere approaches with increasing dimension $\lim\limits_{n\to\infty}{\sigma^2_{B^{(n)}}} \to \frac{1}{2\pi e}$, and for any lattice $\Lambda \subset \R^n$ and all $n$, we have $\sigma_{\Lambda}^2 > \sigma_{B^{(n)}}^2$.
The normalized second moment of a lattice relates the density of the lattice points to the mean square quantization error per dimension, and it was shown by Poltyrev that there exist sequences of lattices $(\Lambda_n \subset \R^n)_n$ which approach
\begin{align*}
	\lim\limits_{n \to \infty}{\sigma^2_{\Lambda_n}} \to \frac{1}{2\pi e},
\end{align*}
a result which can be found in \cite{zamir:nested}. It is thus natural to say that a sequence of lattices is \emph{good for quantization} if it approaches this lower bound in the limit. 

The second quantification of goodness we treat relates to the decoding error probability in an AWGN channel, that is, in a noisy channel without fading. Under this channel model and under the use of lattice codes, ML decoding simply consists of searching for the lattice point closest to the received vector. Consequently, the decoding error probability is precisely given by the probability that the noise shifts the transmitted signal out of its Voronoi cell. More precisely, if $\mb{x} \in \Lambda \subset \R^n$ is the transmitted signal and $\mb{y} = \mb{x} + \mb{n}$ the channel output, the probability of making a decoding error is 
\begin{align*}
	\Pr(\mb{x} \to \hat{\mb{x}}) = \Pr\left(\mb{n} \notin \mc{V}(\Lambda)\right).
\end{align*}
If we denote by $\sigma_n^2(\epsilon)$ the noise variance that attains $\Pr\left(\mb{n}\notin \mc{V}(\Lambda)\right) = \epsilon$, we can define the \emph{volume-to-noise ratio} 
\begin{align*}
	\mu(\Lambda,\epsilon) := \frac{\vl{\Lambda}^{\frac{2}{n}}}{\sigma_n^2(\epsilon)}.
\end{align*}
Similarly to above,  for any $\epsilon \in (0,1)$ the volume-to-noise ratio $\mu(B^{(n)},\epsilon)$ of an $n$-sphere approaches $\lim\limits_{n\to \infty}{\mu(B^{(n)},\epsilon)} \to 2\pi e$ as the dimension grows, and for any lattice $\Lambda \subset \R^n$ and all $n$, we have $\mu(\Lambda,\epsilon) > \mu(B^{(n)},\epsilon)$. A sequence of lattices $(\Lambda_n \subset \R^n)_n$ whose volume-to-noise ratio converges towards this lower limit,
\begin{align*}
	\lim\limits_{n\to\infty}{\mu(\Lambda_n,\epsilon)} \to 2\pi e,
\end{align*}
is called \emph{good for AWGN coding}. The existence of such sequences has been shown by Poltyrev in \cite{poltyrev:nested}.

\chapter{Lattice-Based Communications}
\label{chp:communications}

In this final chapter, we devote our attention to the articles composing this thesis. We classify our work into four different communication settings. Each setup is described in detail in a separate section, wherein we furthermore elucidate the main goals and results of the related publications.

\section{Asymmetric MIMO Channels}
\label{sec:mimo}

We begin with a simple point-to-point communication setup, where the transmitter and receiver are equipped with multiple antennas. In this MIMO setting, the system is called \emph{symmetric}, if the number of transmit and receive antennas coincide, $n_t = n_r$. In contrast, a system where $n_r \neq n_t$ is called \emph{asymmetric}. As mentioned previously, full lattices can be employed for this symmetric setting, and corresponding codes can be (efficiently) decoded simply via an ML procedure. The same codes can also be employed when $n_r > n_t$. However, symmetric codes cannot be optimally decoded if $n_r < n_t$, an assumption that is realistic in many practical scenarios. 

There are various ways of adapting regular symmetric space--time codes to the asymmetric scenario, the most straightforward of which probably is the \emph{block diagonal} construction, a method described \emph{e.g.}, in \cite{hollanti:asymmetric}. We quickly recall this method.

Consider an asymmetric MIMO channel with $n_r$ receive and $n_t = n_r n$ transmit antennas, $n \ge 2$, and let $F \subset K \subset L$ be a tower of cyclic number field extensions with extension degrees $[L:K] = n_r$, $[K:F] = n$, and $[L:F] = n_t = n_rn$. We fix generators of the cyclic Galois groups $\gal(L/F) = \langle \tau \rangle $ and $\gal(L/K) = \langle \sigma \rangle = \langle \tau^n \rangle$, and consider the cyclic division algebra $\mc{C} = (L/K,\sigma,\gamma)$, where $\gamma \in \mc{O}_K\backslash\left\{0\right\}$ is a fixed non-norm element.

Given any order $\Gamma$ in $\mc{C}$, let $\rho(\Gamma)$ be the left-regular representation of the order. Provided that the base field is either $F = \Q$ or quadratic imaginary, the block-diagonal lattice
 \begin{align*}
	\mathcal{L}(\Gamma) = \left\{\left.\begin{bmatrix} \rho(c) & 0 & \cdots & 0 \\ 0 & \tau\left(\rho(c)\right) & & 0 \\ \vdots & & \ddots & \vdots \\ 0 & \cdots & 0 & \tau^{n-1}\left(\rho(c)\right) \end{bmatrix} \in \mat(n_t,\C)\, \right|\, c \in \Gamma \right\}
\end{align*} 
achieves the non-vanishing determinant property. 

The \emph{code rate} of a space--time code carved out from such a block-diagonal lattice $\mc{L}(\Gamma)$ in real symbols per channel use is 
\begin{align*}
	 \mc{R} = \begin{cases}
	 	2n n_r^2/n n_r = 2n_r &\mbox{ if the base field is quadratic imaginary}, \\
 	    2n n_r^2/2n n_r = n_r &\mbox{ if the base field is } \Q. 
 	\end{cases}
 \end{align*}
We remark that while we give the rate in real symbols, it is customary in literature to consider the rate in complex symbols.  

If $\mc{X} \subset \rho(\Gamma)$ is a full-diversity space--time code with non-vanishing determinants, and $F$ is as above, both properties are inherited by a corresponding space--time code in $\mc{L}(\Gamma)$. Motivated by the discriminant minimization problem, or equivalently the maximization of the code density and thus coding gain, in Publication I we are interested in finding explicit orders with smallest possible discriminants for the asymmetric setting. As previously mentioned, among all orders of a cyclic division algebra, maximal orders exhibit the smallest discriminant, and would thus be excellent candidates for code construction. Unfortunately, they are in general very difficult to compute and may result in highly skewed lattices, making the bit labeling a delicate and complex problem on its own. As a compromise between the good performance but complexity of maximal orders, and the simplicity but degraded performance of orthogonal lattices, we examine the attainable lower bound on the discriminants of natural orders, which due to their simpler structure have become a more frequent choice for code construction. 

In Publication I, we fix the extension degree $n \in \left\{1,2,3\right\}$ and pairs of antennas $(n_t,n_r) \in \left\{(2,2),(4,2),(6,2),(6,3)\right\}$, as well as the base field $F = \Q$ or $F = \Q(i)$. The considered settings are interesting as they constitute the asymmetric scenarios which can be found in practice, as modern devices do not allow for much larger numbers of antennas. For each of the considered cases, we give a lower bound on the norm of the discriminant ideal of the natural order $\Gamma_{\mathrm{nat}}$ of any cyclic division algebra respecting the fixed parameters, viewed as an $\mc{O}_F$-module. Furthermore, we give an explicit cyclic number field extension $L/K$ as well as a non-norm element $\gamma \in \mc{O}_K\backslash\left\{0\right\}$, such that the cyclic algebra $(L/K,\sigma,\gamma)$ is division, and its natural order attains the lower bound. 

\section{Amplify-and-Forward Relaying}
\label{sec:AaF}

In a distributed communication setup, such as a wireless relay network, diversity can be enabled via user cooperation. Space--time codes designed to exploit this type of diversity are known as \emph{distributed} space--time codes. In this section, we consider the communication of $(M+1)$ users with a single destination, where every user as well as the destination can be equipped with either a single antenna or multiple antennas. In this scenario, enabling cooperation and dividing the allocated transmission time allows for the $M$ inactive users to aid the active source in communicating with the destination by acting as intermediate \emph{relays}. Equivalently, we can interpret this model as a single-user single-destination communication process with $M$ intermediate relays, which alternate to act as the helping relay. Both interpretations allow for the same code design, and we will adopt both viewpoints interchangeably. The relays are not assumed to have much computational power, as they are not required to perform any kind of decoding. Hence, no channel knowledge is available at the relays, either. Instead, each relay simply amplifies its channel output and forwards it to the destination. The construction of lattice codes for this relaying technique, known as amplify-and-forward relaying, is the subject of Publications II and III. We remark that the notation employed in this section differs from the notation used in the related publications for maintaining consistency within this thesis.

We adopt the assumption that the source and the active relay can transmit information at the same time. This \emph{non-orthogonal} scheme was introduced in \cite{nabar:relay} for single-antenna receivers and sources. The relays are furthermore assumed to operate in \emph{half-duplex} mode, that is, they can only receive or transmit information, but cannot do both simultaneously. 

A generalization of this scheme to the MIMO setting for an arbitrary number of antennas and relays was proposed in \cite{yang:af}, which shall serve as the main reference for this section. In the same article, code criteria were derived for attaining the diversity-multiplexing trade-off of the channel. This MIMO scheme is illustrated in Figure~\ref{fig:aaf_system}. 
\begin{figure}[ht]
\centering
\begin{framed}
\scalebox{.68}{
\begin{tikzpicture}
	\node[relay] (r1) {R$_1$};
	\node[blank,below=0.15 of r1] (vert) {$\vdots$};
	\node[relay,below=0.2 of vert] (rm) {R$_M$};
	\node[blank,left=.4cm of vert] (dummy1) {};
	\node[blank,right=2.5cm of vert] (dummy2) {};
	
	\node[user, above=1.3cm of dummy1] (s) {T};
	\node[user, above=1.3cm of dummy2] (d) {D};
	
	\node[blank_short,right=0.05cm of r1]    (out1) {$\longrightarrow Y_{1,1}$};
	\node[blank_short,right=0.05cm of rm]    (out2) {$\longrightarrow Y_{M,1}$};

	\draw[pil_rev] (d.west) -- (s.east) node[pos=0.5,above]{\scriptsize $H_D$};
	\draw[pil] (s.east) -- (r1.west) node[pos=0.7,above]{\scriptsize $H_{R_1}$};
	\draw[pil] (s.east) -- (rm.west) node[pos=0.7,above,yshift=0.2cm]{\scriptsize $H_{R_M}$};

	\draw[pil_rev] (d.west) -- (out1.east) node[pos=0.7,above]{\scriptsize $H_{D_1}$};
	\draw[pil_rev] (d.west) -- (out2.east) node[pos=0.7,above,yshift=0.2cm]{\scriptsize $H_{D_M}$};

	\draw[black,thick,radiation,decoration={angle=45}] ([xshift=.5cm]s.north) -- +(45:0.5);
	\draw[black,thick,radiation,decoration={angle=45}] ([xshift=.5cm]r1.north) -- +(45:0.5);
	\draw[black,thick,radiation,decoration={angle=45}] ([xshift=.5cm]rm.north) -- +(45:0.5);
	
	\node[blank,left=0.05cm of s] (x) {$\underset{\in \mc{X} \subset \mat(nM,\C)}{\diag\left(X_m\right)_m} \longrightarrow$};

	\node[above=0.02 of d, xshift=1cm] (nd) {\quad\ ${\Big\downarrow} \oplus \left\{N_{i,j}\right\}$};
	\node[right=0.05 of d] (output) {$\underset{\leadsto \left\{X_{1,1},\ldots,X_{M,2}\right\}}{\longrightarrow \left\{Y_{1,1},Y_{1,2},\ldots,Y_{M,2}\right\}}$};
\end{tikzpicture}}
\caption{System model with a single source and destination, and $M$ intermediate relays in half-duplex mode.}
\label{fig:aaf_system}
\end{framed}
\end{figure}

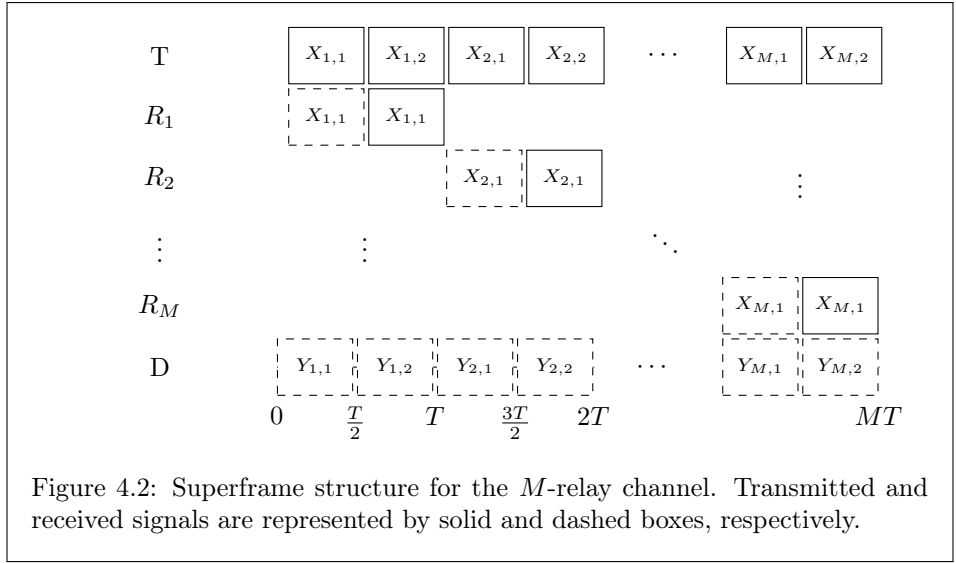
\begin{figure}[ht]
\centering
\begin{framed}
\begin{tikzpicture}

\node[blank]			(source)	                    		{T};
\node[blank]			(r1)		[below=0.1cm of source]		{$R_1$};
\node[blank]    		(r2)    	[below=0.1cm of r1] 		{$R_2$};
\node[blank]    		(dummyr)	[below=0.1cm of r2] 		{$\vdots$};
\node[blank]    		(rN)    	[below=0.1cm of dummyr] 	{$R_M$};
\node[blank]			(dest)		[below=0.1cm of rN]			{D};

\node[blank]			(vert1)		[right=0.01 of dummyr,xshift=-0.3cm] 	{$\vdots$};
\node[blank]			(vert2)		[right=5.5cm of r2]					 	{$\vdots$};

\node[solidnode]		(x11)	[right=0.2cm of source]	{\scriptsize $X_{1,1}$};
\node[solidnode]		(x12)	[right=0.05cm of x11]	{\scriptsize $X_{1,2}$};
\node[solidnode]		(x21)	[right=0.05cm of x12]	{\scriptsize $X_{2,1}$};
\node[solidnode]		(x22)	[right=0.05cm of x21]	{\scriptsize $X_{2,2}$};
\node[dummy]			(dummyt)[right=0.05cm of x22]	{$\cdots$};
\node[solidnode]		(xN1)	[right=0.05cm of dummyt]{\scriptsize $X_{M,1}$};
\node[solidnode]		(xN2)	[right=0.05cm of xN1]	{\scriptsize $X_{M,2}$};

\node[dashednode]		(r11)	[right=0.2cm of r1]		{\scriptsize $X_{1,1}$};
\node[solidnode]		(r12)	[right=0.05cm of r11]	{\scriptsize $X_{1,1}$};

\node[dashednode] 		(r21)	[right=2.3cm of r2]		{\scriptsize $X_{2,1}$};
\node[solidnode] 		(r22)	[right=0.05cm of r21]	{\scriptsize $X_{2,1}$};

\node[dummy]			(dummyd)[below=1.25cm of dummyt,yshift=-0.4cm]{$\ddots$};

\node[dashednode]		(rN1)	[right=5.97cm of rN]	{\scriptsize $X_{M,1}$};
\node[solidnode]		(rN2)	[right=0.05cm of rN1]	{\scriptsize $X_{M,1}$};

\node[dashednode]		(d11) 	[right=0.05cm of dest]	{\scriptsize $Y_{1,1}$}
	edge node[pos=0.5, below=0.4cm]{$0$} (d11.west); 
\node[dashednode]		(d12) 	[right=0.05cm of d11]	{\scriptsize $Y_{1,2}$}
	edge node[pos=0.5, below=0.4cm]{$\frac{T}{2}$} (d11.east); 
\node[dashednode]		(d21) 	[right=0.05cm of d12]	{\scriptsize $Y_{2,1}$}
	edge node[pos=0.5, below=0.4cm]{$T$} (d12.east); 
\node[dashednode]		(d22) 	[right=0.05cm of d21]	{\scriptsize $Y_{2,2}$}
	edge node[pos=0.5, below=0.4cm]{$\frac{3T}{2}$} (d21.east)
	edge node[pos=0.5, below=0.4cm]{$2T$} (d22.east); 
\node[dummy]			(dummyc)	[right=0.05cm of d22]	{$\cdots$};
\node[dashednode]		(dN1) 	[right=0.15cm of dummyc]	{\scriptsize $Y_{M,1}$};
\node[dashednode]		(dN2) 	[right=0.05cm of dN1]		{\scriptsize $Y_{M,2}$}
	edge node[pos=0.5, below=0.4cm]{$MT$} (dN2.east); 
\end{tikzpicture}
\caption{Superframe structure for the $M$-relay channel. Transmitted and received signals are represented by solid and dashed boxes, respectively.}
\label{fig:aaf_frame}
\end{framed}
\end{figure}

The matrices $H_D$, $H_{R_m}$ and $H_{D_m}$, $1 \le m \le M$ denote the Rayleigh distributed channels from the source to the destination, relays, and from the relays to the destination, respectively. In a first time slot, the source communicates simultaneously with the first relay, as well as directly with the destination. The relay amplifies its channel output and, in a second time slot, forwards this amplified signal to the destination. During this second time slot, the source also communicates directly to the destination. This process is repeated with each of the relays. To illustrate this layered process, define a superframe consisting of $M$ consecutive cooperation frames, during which the relays take turns to cooperate with the active source. Each frame of length $T$ is composed of two partitions of $T/2$ symbols. This frame model is depicted in Figure~\ref{fig:aaf_frame}. 

Denote by $n_t$ the number of antennas at the source, and let $n_r$ denote the number of receive antennas at the destination. Further, let each relay be equipped with $n_R \le n_t$ antennas. For each cooperation frame, the output at the destination for each half of the frame is given by
\begin{align*}
	Y_{m,1} &= \rho_1 H_D X_{m,1} + N_{m,1}, \\
	Y_{m,2} &= \rho_2 H_D X_{m,2} + N_{m,2} + \rho_3 H_{D_m} B_m \tilde{X}_{m,1},
\end{align*}
where $\tilde{X}_{m,1} = \rho_1' H_{R_m} X_{m,1} + N_{m}'$ is the channel output at the relay for the first half of the frame. The matrices $N_{m,i}$ and $N_{m}'$ represent additive white Gaussian noise, the matrices $B_m$ are used for amplification and $\rho, \rho'$ are power allocation factors. 
 
From the destination's point of view, we can equivalently present this communication process as a virtual single-user MIMO channel model. Setting $T = n := M(n_t+n_R)$, we get the familiar channel equation $Y = HX + N$,
where $X \in \mat(n,\C)$ and $Y \in \mat(n_r\times n,\C)$ are the (overall) transmitted and received signals, and the structure of the channel matrix $H \in \mat(n_r\times n,\C)$ is determined by the different relay paths. 

Consider a block-diagonal space--time code $\mathcal{X}$, that is, where each $X \in \mathcal{X}$ takes the form 
\begin{align*}
	X = \diag\left(X_m\right)_m = \begin{bmatrix} X_1 & & \\ & \ddots & \\ & & X_M \end{bmatrix}
\end{align*}
with $X_m \in \mat(2n_t, \mathbb{C})$. It was shown in \cite{yang:af} that such codes which additionally respect the usual design criteria, such as non-vanishing determinants, are good choices for this channel. 

Our interest in Publications II and III is to give explicit construction methods for distributed space--time codes that can be employed in this MIMO setting, but are additionally fast-decodable. To adapt the codes to suit the considered setting, we make use of an iterative construction proposed in \cite{markin:iterated}, which modifies the underling algebra and, under certain conditions, ensures that the iterated codes inherit certain properties. To briefly summarize the iterative construction, consider a cyclic division algebra $\mc{C} = (L/K,\sigma,\gamma)$ of degree $n$, where $K$ is a finite Galois extension of $\Q$. For $\theta = \zeta\theta' \in \mc{C}$ fixed with $\zeta \in \left\{\pm 1, \pm i \right\}$ and $\theta' \in \R_{>0}$, and for $\tau$ a $\Q$-automorphism of $L$, we define the function
\begin{align*}
	\tilde{\alpha}_{\tau,\theta}: \mat(n,L)\times\mat(n,L) &\to \mat(2n,L) \\
	(X,Y) &\mapsto \begin{bmatrix} X & \zeta\sqrt{\theta'}\tau(Y) \\ \sqrt{\theta'}Y & \tau(X) \end{bmatrix}.
\end{align*}
Here, the function $\tau$ is applied to each of the coefficients of the matrices $X$ and $Y$. Suppose that the algebra $\mc{C}$ gives rise to a rank-$k$ space--time code $\mc{X}$ defined by the matrices $\left\{ B_i \right\}_{i=1}^{k}$. Then, for a signaling alphabet $S$, the matrices $\left\{ \tilde{\alpha}_{\tau,\theta}(B_i,0), \tilde{\alpha}_{\tau,\theta}(0,B_i)\right\}_{i=1}^{k}$ define a rank-$2k$ code 
\begin{align*}
	\mc{X}_{\mathrm{it}} = \left\{\left. \sum\limits_{i=1}^{k}\left[\tilde{\alpha}_{\tau,\theta}(B_i,0)s_i + \tilde{\alpha}_{\tau,\theta}(0,B_i)s_{k+i}\right] \right| s_i \in S\right\}. 	
\end{align*}

Under carefully ensured conditioned, the code $\mc{X}_{\mathrm{it}}$ retains both the full-diversity and the non-vanishing determinants property. Furthermore, if for some $i,j$ we have $B_iB_j^\dagger + B_jB_i^\dagger = 0$, then 
\begin{align*}
	\tilde{\alpha}_{\tau,\theta}(B_i,0)\tilde{\alpha}_{\tau,\theta}(B_j,0)^\dagger + \tilde{\alpha}_{\tau,\theta}(B_j,0)\tilde{\alpha}_{\tau,\theta}(B_i,0)^\dagger &= 0,\\
	\tilde{\alpha}_{\tau,\theta}(0,B_i)\tilde{\alpha}_{\tau,\theta}(0,B_j)^\dagger + \tilde{\alpha}_{\tau,\theta}(0,B_j)\tilde{\alpha}_{\tau,\theta}(0,B_i)^\dagger &= 0.
\end{align*}

The first question that we pose is whether using this construction and imposing the fast-decodability property results in degraded performance of the codes, a problem which we study empirically in Publication II. Therein, we consider algebraic codes which are known to perform well, such as the Silver \cite{paredes:silver} and Golden code \cite{belfiore:golden}, as well as a code constructed in \cite{vehkalahti:asymmetric} from cyclotomic extensions. The former two codes need to be first adapted to the specific channel setting to fit the chosen number of antennas, to which end we make use of the aforementioned iterative method. Further, the resulting iterated codes are diagonalized employing a suitable map of order $M$. For a fixed number of relays, the observation is that, even after iterating the underlying algebra, the performance of the resulting codes does not suffer from the additionally forced properties. 

The good performance of the example codes constructed in Publication II motivates further study of general constructions of families of fast-decodable distributed space--time codes. This is the topic of Publication III. In the first step, we investigate the case $(n_t,n_R) = (1,1)$ and $n_r \ge 2$. Preliminary results for this scenario have been presented in \cite{hollanti:relay1, hollanti:relay2}. Our construction relies on cleverly chosen field extensions and related cyclic division algebras, and we again make use of the iterative construction and suitable diagonalization function. The first result gives raise to an infinite family of distributed space--time codes with non-vanishing determinants, which have full rate for $n_r = 2$. In addition, the codes are conditionally $4$-group decodable, and as such achieve a reduction in decoding complexity of $37.5\%$ in contrast to non-fast-decodable space--time codes of equal rank.

In a second step, we generalize the setting and allow for multiple antennas at the source and relays, and assume $n_r \ge 1$ antennas at the destination. Our construction for this more general setting makes use of the maximal real subfield of a cyclotomic field, which relates to the number of relays considered. Hence, the number of relays needs to be expressible as $M =  (p-1)/2$ for $p \ge 5$ prime. We again make use of the iterative construction in order to adapt the underlying algebraic structure to the relay channel, and then diagonalize the resulting codes employing a Galois automorphism of suitable degree. The proposed construction gives rise to an infinite family of distributed space--time codes with non-vanishing determinants, which have full rate for a single receive antenna at the destination. Remarkably, the codes arising from this construction are either $2$- or $4$-group decodable, resulting in a reduction in decoding complexity of $50\%$ and $75\%$, respectively. These are the first constructions of fast-decodable space--time codes for the MIMO amplify-and-forward channel. It is noteworthy that according to recent work \cite{berhuy:fd_bounds}, group decodable codes from cyclic division algebras cannot exceed $g = 4$ groups, thus $75\%$ is the maximum possible reduction in decoding complexity.

Additionally, we construct various examples of codes using the proposed results, giving \emph{i.a.}, the first explicit examples for $M \ge 3$ that can be found in the literature, and compare their performance to other well-performing codes lacking the fast-decodability property.

We conclude the article with an adaptation of the constructions to the multiple-access channel, and we show that it is straightforward to construct fast-decodable codes for this scenario using the presented methods.

\section{Compute-and-Forward Relaying}
\label{sec:CaF}

In the relaying technique introduced in the previous section, intermediate relays cause delays due to the half-duplex assumption. Imitating the notion of \emph{network coding}, \emph{i.e.}, a technique employed at the network layer where intermediate nodes compute functions of incoming packets, which are then forwarded across the network, \emph{physical layer network coding} pursues a similar strategy at a different level of communication. A particularly promising protocol for physical layer network coding has been introduced by Nazer and Gastpar in the award winning paper \cite{nazer:cf}. Operating under this \emph{compute-and-forward} protocol, sources employ nested lattice codes and transmit single codewords to multiple intermediate relays. Each relay observes a noisy superposition of the incoming codewords, and attempts to decode the channel output to a linear combination of the received messages. The linear combinations are then forwarded to an ultimate destination or possibly to further relays. This circumvents decoding the individual messages at the relays and hence potentially increases the throughput.

The compute-and-forward protocol, which will be introduced more carefully in what follows, is central to the work carried out in Publications IV and V. The original paper \cite{nazer:cf} serves as our main reference.

Consider $K > 1$ sources, communicating with a single destination aided by $M$ intermediate relays. We assume that each source, relay, and destination is equipped with one antenna only. The first hop from the sources to the relays is modeled as a Gaussian fading channel, while it is typically assumed that the relays are connected to a destination with error-free bit pipes. As we are only considering real-valued channels, we present the protocol in its real version. Complex channels can be discussed similarly, after transforming the complex channel output into two equivalent real channel outputs that can be treated separately \cite{nazer:cf}. 

The sources want to communicate messages $\mb{w}_k \in \F_p^s$ to the destination. Before transmission, these messages are encoded into $n$-dimensional codewords, $\mb{w}_k \mapsto \mb{x}_k \in \Lambda_{k,F} \subset \R^n$, where $\Lambda_{k,F}$ is a full lattice employed by source $k$, acting as the fine lattice in the nested code $\mc{C}_k(\Lambda_{k,C},\Lambda_{k,F}) = \left\{\left[\mb{x}\right] \in \Lambda_{k,F} \left(\bmod\ \Lambda_{k,C} \right) \mid \mb{x} \in \Lambda_{k,F} \right\}$. We impose the usual symmetric power constraint $\frac{1}{n} \mathbb{E}\left[\|\mb{x}_k\|^2\right] \leq P$ for all $k$. 

Each relay is only assumed to have information about the channel to itself, and observes a noisy superposition of the transmitted codewords, that is, the channel output at relay $m$ is
\begin{align*}
	\mb{y}_m = \sum\limits_{k=1}^{K}{h_{mk}\mb{x}_k} + \mb{n}_m.
\end{align*}

In contrast to the well-known \emph{decode-and-forward} protocol, the goal of the relay is not to estimate the individual codewords $\mb{x}_k$. Instead, it attempts to compute a linear combination of the transmitted codewords, \emph{i.e.}, given a coefficient vector $\mb{a}_m^t = (a_{m1},\ldots,a_{mK})$, it estimates 
\begin{align*}
	\lambda_m = \sum\limits_{k=1}^{K}{a_{mk}\mb{x}_k},
\end{align*}
and forwards this linear combination to the destination. The destination, given the coefficient matrix $A = (\mb{a}_k)^t_k$ as well as the linear combinations, attempts to solve for the original messages. The model is depicted in Figure~\ref{fig:caf_system}.
\begin{figure}[!h]
\centering
\begin{framed}
\scalebox{.68}{
\begin{tikzpicture}
	\node[relay] (r1) {R$_1$};
	\node[blank,below=0.2 of r1] (vert2) {$\vdots$};
	\node[relay,below=0.3 of vert2] (rm) {R$_M$};
	\node[blank,left=.7 of vert2] (vert1) {$\vdots$};
	
	\node[user, above=0.2 of vert1] (t1) {T$_1$};
	\draw[pil] (t1.east) -- (r1.west) node[pos=0.4,above,xshift=0.1cm]{\scriptsize $h_{11}$};
	\draw[pil] (t1.east) -- (rm.west) node[pos=0.3,above,xshift=0.1cm]{\scriptsize $h_{M1}$};
	
	\node[user, below=0.3 of vert1] (tk) {T$_K$};
	\draw[pil] (tk.east) -- (r1.west) node[pos=0.3,above,xshift=0.1cm,yshift=0.1cm]{\scriptsize $h_{1K}$};
	\draw[pil] (tk.east) -- (rm.west) node[pos=0.4,above,xshift=0.1cm]{\scriptsize $h_{MK}$};
		
	\node[blank,right=.1 of r1] (lincomb1) {$\longrightarrow\ \mb{y}_1 \leadsto \lambda_1$};
	\node[blank,right=.1 of rm] (lincombm) {$\longrightarrow\ \mb{y}_M  \leadsto \lambda_M$};
	
	\node[user, right=3.75 of vert2] (dest) {D};
	\draw[pil_rev] (dest.west) -- (lincomb1.east);
	\draw[pil_rev] (dest.west) -- (lincombm.east);
		
	\node[above=0.02 of r1, xshift=.7cm] (n1) {\quad\ ${\Big\downarrow} \oplus \mb{n}_1$};
	\node[above=0.02 of rm, xshift=.7cm] (nm) {\quad\ ${\Big\downarrow} \oplus \mb{n}_M$};
	\node[above=0.02 of dest, xshift=.3cm] (A) {\quad\ $\Big\downarrow A$};
	
	\node[blank, right=.1 of dest] (cwords) {$\leadsto \left\{\mb{x}_1,\ldots,\mb{x}_K\right\}$};	
	\node[blank,left=.2 of t1] (x1) {$\underset{\in \F_q^k}{\omega_1} \longmapsto \underset{\in \Lambda_{1,F}/\Lambda_{1,C}}{\mb{x}_1} \longrightarrow$};
	\node[blank,left=.2 of tk] (xl) {$\underset{\in \F_q^k}{\omega_K} \longmapsto \underset{\in \Lambda_{K,F}/\Lambda_{K,C}}{\mb{x}_K} \longrightarrow$};
	
	\draw[black,thick,radiation,decoration={angle=45}] ([xshift=.5cm]t1.north) -- +(45:0.5);
	\draw[black,thick,radiation,decoration={angle=45}] ([xshift=.5cm]tk.north) -- +(45:0.5);

\end{tikzpicture}}
\caption{System model with $K > 1$ sources and $M > K$ relays connected to a destination.}
\label{fig:caf_system}
\end{framed}
\end{figure}

The remainder of this section is split into two parts. We first consider the hop from the sources to the intermediate relays, introduce quantities in this context measuring the performance of the compute-and-forward scheme, and explain how the relays can compute the target linear combinations. We then study the hop from the relays to the destination.

\subsection*{From the Sources to the Relays}
\label{subsubsec:hop1}

As presented above, assume $K \ge 2$ sources and $M \ge K$ intermediate relays. The first important metric for performance analysis in compute-and-forward is the \emph{computation rate} $\mc{R}_C(\mb{a},\mb{h})$, which imposes an upper bound on the code rate at the sources. More specifically, let each source $1 \le k \le K$ employ a nested lattice code with rate $\mc{R}_k$. A relay can decode a linear combination involving the messages whose corresponding rate does not exceed the computation rate achieved at the relay. If we denote the $\snr$ by $\rho = P/\sigma_n^2$, the achievable computation rate region at the $m^{\text{th}}$ relay is given by 
\begin{align*}
	\mc{R}_C(\mb{a}_m,\mb{h}_m) = \max\limits_{\alpha_m \in \R}{\frac{1}{2}\log^+\left(\frac{\rho}{\alpha_m^2 + \rho\|\alpha_m\mb{h}_m - \mb{a}_m\|^2}\right)},
\end{align*}
and the relay is able to decode a linear combination of codewords with rate $\mc{R}_k \le \mc{R}_C(\mb{a}_m,\mb{h}_m)$. By solving a minimum mean square error problem, it can be shown that for a fixed channel and coefficient vector $(\mb{h}_m$, $\mb{a}_m)$, the computation rate is maximized for the specific choice $\alpha_m = \frac{\rho\mb{h}_m^t\mb{a}_m}{1+\rho\|\mb{h}_m\|}$, which results in the computation rate region 
\begin{align*}
	\mc{R}_C(\mb{a}_m,\mb{h}_m) = \frac{1}{2}\log^+\left(\left(\|\mb{a}_m\|^2 - \frac{\rho(\mb{h}_m^t \mb{a}_m)^2}{1+\rho\|\mb{h}_m\|^2}\right)^{-1}\right).
\end{align*}

One of the main results in \cite{nazer:cf} provides design criteria for the lattices $\Lambda_{k,C}$ and $\Lambda_{k,F}$. More specifically, it is shown that if $\Lambda_{k,C} = \Lambda_C$ is a common superlattice for each source and the fine lattices are nested, $\Lambda_{1,F} \supseteq \cdots \supseteq \Lambda_{K,F}$, then for all channel vectors and for all coefficient vectors, relay $m$ can decode the linear combination $\lambda_m$ with the given coefficients with vanishing error probability, provided that the code rates do not exceed the instantaneous computation rate. There are two crucial properties in the proof of this statement, namely
\begin{itemize}
	\item[i)] the coarse lattice $\Lambda_C$ should be good for quantization, 
	\item[ii)] the fine lattices $\Lambda_{1,F},\ldots,\Lambda_{M,F}$ should be good for AWGN coding.
\end{itemize} 

Having computed the maximum achievable computation rate, the next problem posed is the choice of coefficient vector determining the target linear combination. This is solved in \cite{osmane:caf_implementation}, wherein it is shown that the optimal coefficient vector maximizing the computation rate is the solution to the minimization problem
\begin{align}
\label{eqn:svp}
	\mb{a}_{\mathrm{opt}} = \argmin\limits_{\mb{a} \in \Z^K\backslash\left\{\mb{0}\right\}}{\mb{a}^tG\mb{a}},
\end{align}
where $G = I_K-\frac{\rho\mb{h}_m\mb{h}_m^t}{1+\rho\|\mb{h}_m\|^2}$, and this minimization problem corresponds to the shortest vector problem in the lattice with Gram matrix $G$. 

We now briefly describe two methods for the relays to compute the desired linear combination. Note that after observing the channel output, each relay proceeds in the same fashion. We henceforth drop the subscript related to the relay for notational ease. Note further that for any coefficient vector $\mb{a}$, the linear combination $\lambda_m$ is an element in the lattice 
\begin{align*}
	\Lambda_F := \sum\limits_{k=1}^{K}{a_k \Lambda_{k,F}}.
\end{align*}
If $\mb{a}$ is the solution to \eqref{eqn:svp}, then $\gcd(a_i) = 1$ and as the fine lattices are nested, we have $\Lambda_F = \Lambda_{k_{\min},F}$, where $k_{\min}$ is the first non-zero coefficient of $\mb{a}$. We write $\Lambda_F$ for the lattice in which the desired linear combination lives.

\subsubsection{Shortest Distance Decoding}
\label{subsubsec:short_decoder}

Assume that a fixed relay observes the channel output $\mb{y}$ as described above. As it has information about the channel to itself, it solves the minimization problem \eqref{eqn:svp} to estimate the best coefficient vector, and subsequently computes the optimal scaling factor $\alpha$. The channel output can be scaled and rewritten to read 
\begin{align*}
	\tilde{\mb{y}} = \alpha\mb{y} = \sum\limits_{k=1}^{K}{\alpha h_k \mb{x}_k} + \alpha\mb{n} = \sum\limits_{k=1}^{K}{a_k \mb{x}_k} + \left(\sum\limits_{k=1}^{K}{(\alpha h_k - a_k)\mb{x}_k} + \alpha\mb{n}\right).
\end{align*}

The term $\mb{n}_{\mathrm{eff}} := \sum\limits_{k=1}^{K}{(\alpha h_k - a_k)\mb{x}_k} + \alpha\mb{n}$ is referred to as the \emph{effective noise}, and it is important to note that it is no longer Gaussian.

The relay, however, simply treats the scaled channel output as a Gaussian channel equation $\tilde{\mb{y}} = \lambda + \mb{n}_{\mathrm{eff}}$ and estimates the element in $\Lambda_{F}$ closest to the scaled signal, that is, computes $\hat{\lambda} = \argmin\limits_{\mb{x} \in \Lambda_{F}}\|\tilde{\mb{y}}-\mb{x}\|^2$. 

In a second step, the estimated lattice point is shifted back to the basic Voronoi cell $\mc{V}(\Lambda_C)$ by computing $\left[\hat{\lambda}\right] = \hat{\lambda} \bmod\ \Lambda_C$.

\subsubsection{Maximum Likelihood Decoding}
\label{subsubsec:ml_decoding}

In the context of compute-and-forward, ML decoding amounts to maximizing the conditional probability 
\begin{align*}
	\hat{\lambda} &=  \argmax\limits_{\lambda \in L_F}{\Pr\left(\alpha\mb{y}\mid\lambda\right)} \\
	&= \argmax\limits_{\lambda \in L_F}{\sum\limits_{\substack{(\mb{x}_1,\ldots,\mb{x}_K) \in (\mc{C}_1,\ldots,\mc{C}_K) \\ \sum\limits_{k=1}^K{a_k\mb{x}_k} = \lambda}}}{\Pr\left(\alpha\mb{y}\mid(\mb{x}_1,\ldots,\mb{x}_K)\right)\Pr\left((\mb{x}_1,\ldots,\mb{x}_K)\right)},
\end{align*}
where $L_F \subset \Lambda_F$ is finite, determined by the imposed power constraint and the coefficients of the linear combination. By assuming equiprobable codewords in $(\mc{C}_1,\ldots,\mc{C}_K)$, it can be shown that an estimate $\hat{\lambda}$ can be computed by solving $\hat{\lambda} = \argmax\limits_{\lambda \in L_F}{\varphi(\lambda)}$, where
\begin{align*}
	\varphi(\lambda) := \sum\limits_{\substack{(\mb{x}_1,\ldots,\mb{x}_K)\in (\mc{C}_1,\ldots,\mc{C}_K) \\ \sum\limits_{k=1}^K{a_k\mb{x}_k} = \lambda}}{\exp\left\{-\frac{1}{2\sigma^2}\left|\left|\mb{y}-\sum\limits_{k=1}^K{h_k\mb{x}_k}\right|\right|^2\right\}}.
\end{align*}

In Publication IV, we study the behavior of the ML decoding metric, a topic on which not much work exists. In \cite{belfiore:lattices_caf, belfiore:flatness_caf}, two short publications establishing the foundation for our work, the authors examine the decoding function, relating its behavior to the flatness factor of a certain sum of random lattices, and propose an efficient decoding algorithm in dimension $n = 1$ based on Diophantine approximation. Subsequently, the authors in \cite{mejri:decoding_caf} examine the decoding complexity and performance of said algorithm. They further investigate the decoding problem in Gaussian channels without fading, and propose efficient decoding algorithms for this scenario.

Closely following the articles \cite{belfiore:lattices_caf, belfiore:flatness_caf}, we start in Publication IV with a manipulation of the function $\varphi(\lambda)$, which allows us to express the decoding metric directly in terms of the target linear combination $\lambda$. In contrast to previous work, we allow for arbitrary nested lattices at the sources. The decoding problem is shown to read
\begin{align*}
	\hat{\lambda} = \argmax\limits_{\lambda \in L_F}{\sum\limits_{\mb{t} \in S \subset \Z^{nK}}{\exp\left\{\frac{1}{2\sigma^2}\left\|\omega(\lambda)-M_{\mc{L}}\hat{U}\mb{t}\right\|^2\right\}}},
\end{align*}
where $\omega(\lambda)$ is explicitly given in terms of $\lambda$. The important object in this equation is the matrix $M_{\mc{L}} \in \mat(n\times n(K-1),\R)$, which defines a sum of $K-1$ random lattices $\mc{L}$. Previously, it had been misleadingly assumed that $\mc{L}$ is a lattice for any number of sources, while this is only true for $K = 2$. 

\begin{figure}[h!]
	\begin{subfigure}[t]{0.49\textwidth}
		\includegraphics[width=\textwidth]{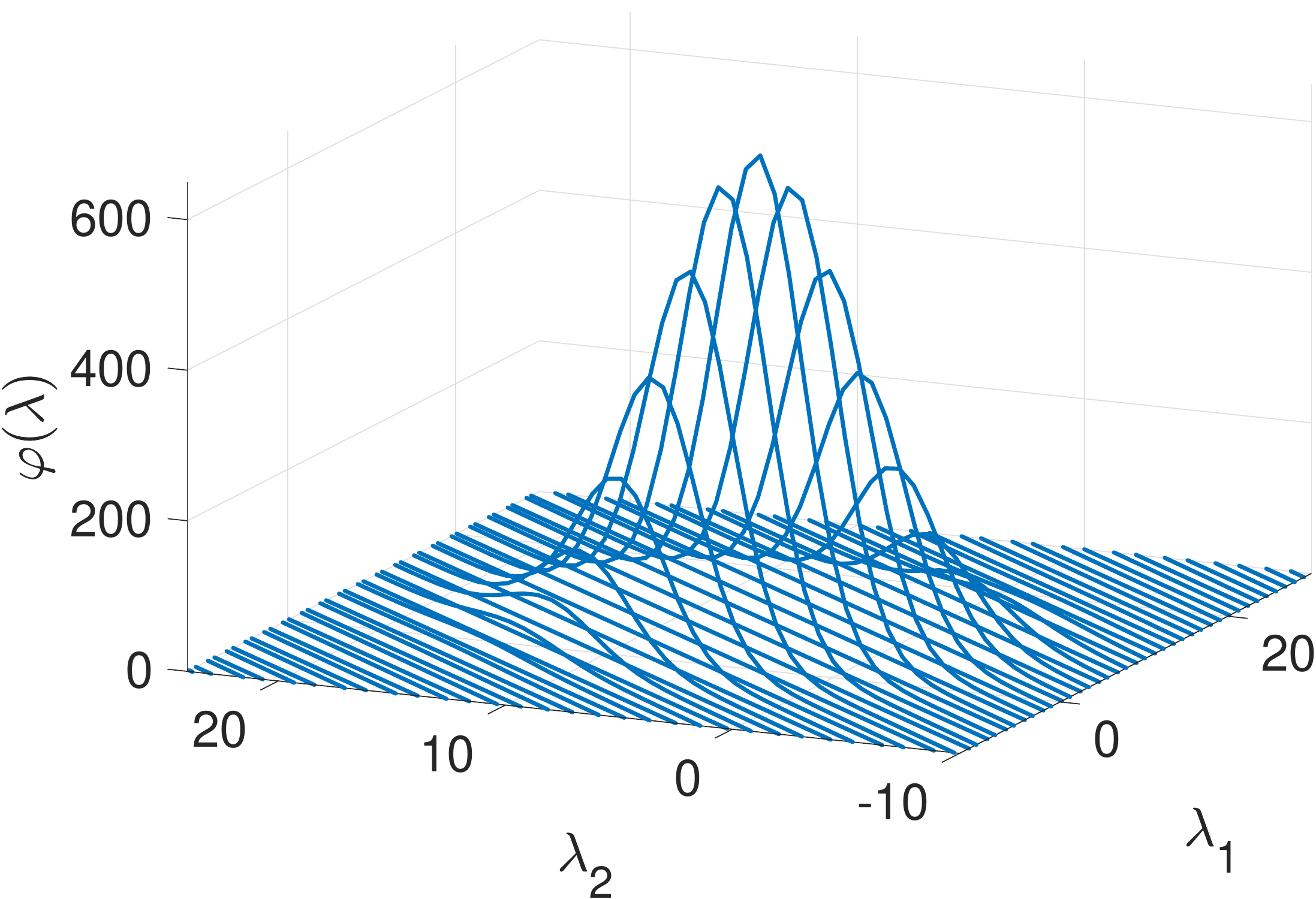}
	\end{subfigure}\hfill
	\begin{subfigure}[t]{0.49\textwidth}
		\includegraphics[width=\textwidth]{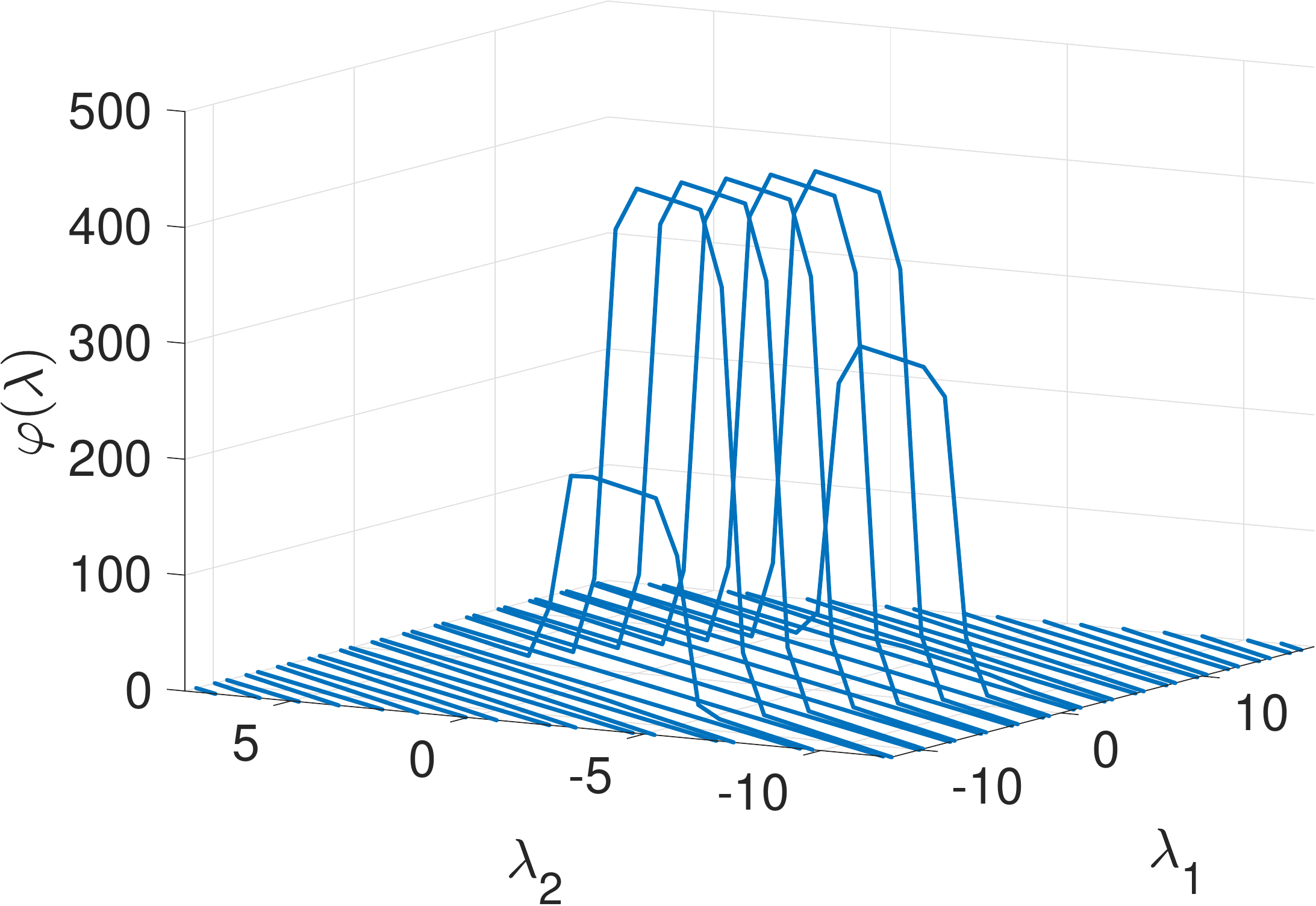}
	\end{subfigure}	
	\caption{Flat behavior of the decoding metric exemplified on the lattice $\Psi\left(\mc{O}_{\Q(\sqrt{5})}\right)$, the lattice constructed from the ring of integers $\mc{O}_K$ of the number field $K = \Q(\sqrt{5})$ via the canonical embedding.}
	\label{fig:caf_flat}
\end{figure}

In \cite{belfiore:lattices_caf, belfiore:flatness_caf}, assuming integer lattices at the sources, it has been shown that the decoding metric can exhibit a flat behavior, which leads to ambiguous decoding decisions and thus errors at the relay. Setting $K = 2$, it was further shown that the flat behavior can be related to the flatness factor of the lattice $\mc{L}$. Following this work, we first show that the flat behavior prevails when relaxing the integer condition. We illustrate this in Figure~\ref{fig:caf_flat}. 
On the left figure, the channel realization and optimal coefficient vector result in a decoding metric which exhibits a unique maximum, and the linear combination computed by the relay is precisely the lattice point $\lambda \in \Lambda_F$ corresponding to this maximum. On the other hand, the right figure depicts an instance where the decoding metric is \emph{flat}, and maximized for multiple values of $\lambda$. This results in ambiguous decisions and, ultimately, errors. 

Adopting the assumptions in \cite{belfiore:lattices_caf, belfiore:flatness_caf}, we show that in order to maximize the flatness factor of the lattice $\mc{L}$, it suffices to maximize that of the code lattice. This is an explicit design criterion for the code lattice, and yields a potential code design trade-off with the usual goodness criteria. In the same article, we derive Theorem~\ref{thm:theta_approx} presented here in Section~\ref{sec:lattices}, a result which we then use to empirically analyze the flatness factor of various lattices.

\subsection*{From the Relays to the Destination}
\label{subsec:hop2}

After decoding a linear combination, each of the relays forwards the estimated lattice point to the destination. The goal of the receiver is to recover the original messages given the following two ingredients: 
\begin{itemize}
	\item[i)] $M$ linear combinations $\lambda_m = \sum\limits_{k=1}^{K}{a_{km}\mb{x}_k}$.
	\item[ii)] The coefficient matrix $A = \left(a_{ij}\right)_{i,j} = \left(\mb{a}_1 \cdots \mb{a}_K \right)^t$.
\end{itemize}

In current research, it is usually assumed that the transmission from the relay to the destination is error-free, and the destination is expected to simply have access to both items without the need to decode. It is obvious that the destination can only recover the original messages if $K \ge M$ and $\rk{A} = M$. However, as the relays compute their optimal coefficient vector independently of each other by solving the shortest vector problem \eqref{eqn:svp}, there is no guarantee that the matrix $A$ should be of full-rank $M$. 

\begin{figure}[h!]
	\includegraphics[width=.75\textwidth]{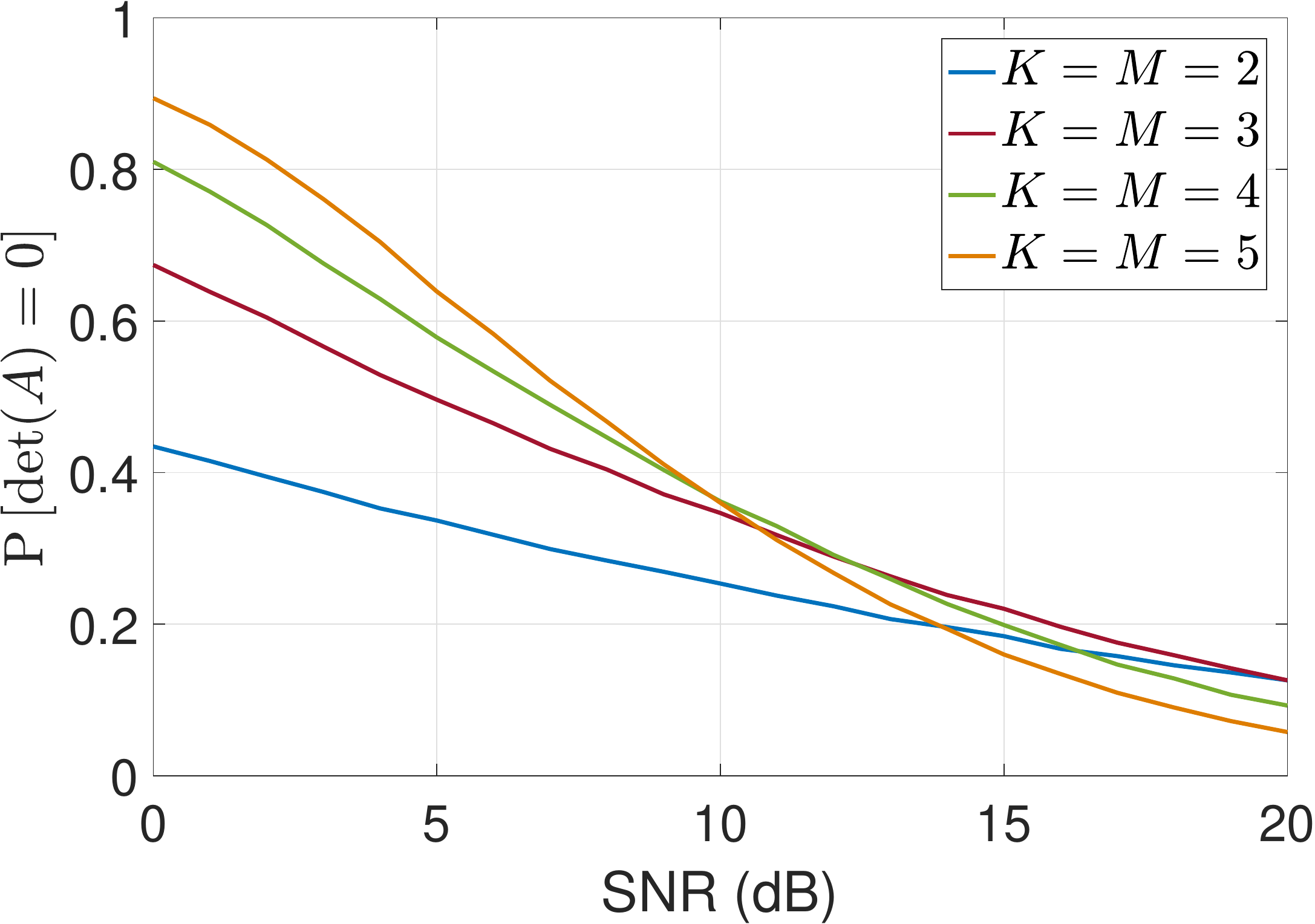}
	\caption{Probability of the equation coefficient matrix $A$ being singular when each relay chooses its optimal coefficient vector.}
		\label{fig:caf_deterror}
\end{figure}

For varying $M = K$, Figure~\ref{fig:caf_deterror} illustrates that the choice of optimal coefficient vectors often result in a non-trivial probability of the matrix $A$ being singular. 

In Publication V we observe that only few coefficient vectors are typically optimal for bounded $\snr$ values. Based on this observation, we propose a method for $M = K = 2$ where we compile small disjoint coefficient vector candidate sets, which are assigned to the relays prior to communication. The splitting of the coefficient vectors is based on the non-degeneracy of the action of a rotation matrix on the quotient $(\Z^2\backslash\left\{(0,0)\right\})$ $\bmod\ \Z^2$, which allows us to divide this space into two disjoint and complementary sets. A relay attempting to decode a linear combination then simply chooses the coefficient vector within its assigned set maximizing the instantaneous computation rate. We furthermore show that, independently of the channel quality, the computation rate achieved using the proposed method is always non-zero, and in expectation both relays achieve the same computation rate. 

The drawback of the proposed method is the in average lower achievable computation rate at each relay in contrast to the optimal method of solving the shortest vector problem. However, the advantage of the method in Publication V is twofold. On one hand, no cooperation is assumed between the relays, and each individual relay only needs to search for a suitable coefficient vector within a small set of candidate vectors. Thus, the complexity of this approach is low. Secondly, it is guaranteed that the system of linear equations is solvable at the destination, which eliminates the need for retransmissions.

\section{Wiretap Coset Codes}
\label{sec:wiretap}

The last communication setup considered in this thesis is the fading \emph{wiretap channel}. First introduced over four decades ago in a wired communications context \cite{wyner:wiretap, ozarov:wiretap}, this model has recently become exceedingly important for wireless networks. Especially the design of suitable lattice codes has received much attention. Nested lattice code design for this particular model is the main subject of Publications VI and VII. 	

In a wiretap channel model, two legitimate communication parties, usually referred to as Alice and Bob, attempt to communicate securely over a wireless channel in the presence of an eavesdropper, Eve. Here, secure exchange of information relies on physical layer security rather than traditional cryptographic protocols, though a combination of the both is naturally encouraged; the message might be encrypted before transmission, a type of processing which occurs at the network layer. We are however interested in providing security at the physical layer. 

Both receivers, Bob and Eve, have perfect channel state information, though Eve's channel is assumed to be of worse quality than Bob's. This is a typical assumption, and can either be naturally satisfied depending on the actual physical environment, or otherwise artificially achieved, \emph{e.g.}, through beamforming or jamming. When designing a code for this wiretap scenario, it is equally important to ensure that Bob can correctly decode Alice's message, while simultaneously lowering Eve's chances of successfully learning any information from her intercepted signal. We are mostly interested in the latter. As in the previous compute-and-forward setting, codes are constructed from nested lattices. 

We allow each communication party to be equipped with multiple antennas. More concretely, let $n_t$, $n_b$ and $n_e$ denote the number of transmit antennas for Alice, and the receive antennas for Bob and Eve, respectively. Coding over $T$ time slots, Alice transmits a codeword $X \in \mat(n_t\times T,\C)$, and the channel outputs of Bob and Eve are given by the equations
\begin{align*}
	Y_b = H_b X + N_b, \quad Y_e = H_e X + N_e.
\end{align*}
Here, $H_b$ and $H_e$ denote the $n_b\times n_t$ and $n_e\times n_t$ channel matrices with i.i.d. complex Gaussian entries, and $N_b$, $N_e$ are the additive white Gaussian noise matrices. 

To focus on the lattice structure of the code, we identify $\mat(n_t\times T,\C)$ with the vector space $\R^{n}$ using the isometry $\iota$ (cf. \eqref{eqn:isometry}), where we define $n := 2n_t T$. We have the equivalent vectorized channel outputs 
\begin{align*}
	\iota(Y_b) &= (I_{T}\otimes \overline{H}_b)\iota(X) + \iota(N_b), \\ 
	\iota(Y_e) &= (I_{T}\otimes \overline{H}_e)\iota(X) + \iota(N_e),
\end{align*}
where $\overline{H} = \left(\overline{h}_{ij}\right)_{i,j}$ with
\begin{align*}
	\overline{h}_{ij} = \begin{bmatrix} \Re(h_{ij}) & -\Im(h_{ij}) \\ \Im(h_{ij}) & \Re(h_{ij}) \end{bmatrix}
\end{align*}
We will henceforth use the notation $X$ or $\mb{x} = \iota(X)$ for the codeword in $\mat(n_t\times T,\C)$ and $\R^n$, respectively, and simply write $\Lambda$ for the lattice in either ambient space, as the context will always be clear. 

Alice is equipped with a pair of nested full lattices $\Lambda_e \subset \Lambda_b \subset \R^n$. Let $\mf{M}$ denote Alice's message set of cardinality $|\mf{M}| = |\Lambda_b/\Lambda_e|$. The original messages are encoded into the set of unique coset representatives of $\Lambda_b/\Lambda_e$ via an injective map
\begin{align*}
	\mc{E}: \mf{M} \to \Lambda_b\cap \mc{V}(\Lambda_e); \quad m \mapsto \mb{x}_m.
\end{align*}  
Thus, $\mb{x}_m$ is a point in the lattice $\Lambda_b$, and contains the information bits intended for the legitimate receiver, Bob. Further, Alice purposefully adds random bits to the message in order to confuse the eavesdropper. She randomly picks an element $\mb{x}_r \in \Lambda_e$ and computes $\mb{x} = \mb{x}_m + \mb{x}_r \in \mb{x}_m+\Lambda_e$. Note that there are various ways of choosing $\mb{x}_r$. Further, using this method the same message is mapped to several different lattice codewords, hence the set of possible codewords is larger than the set of original messages. 

\begin{figure}[ht]
\centering
\begin{framed}
\scalebox{0.97}{
\begin{tikzpicture}
	\node[user] (Bob) {Bob};
	\node[user, left=3.2cm of Bob] (Alice) {Alice};
	\node[user, below=1.5 of Bob, xshift=-1.5cm] (Eve) {Eve};

	\node[above=0.02 of Bob, xshift=.7cm] (zb) {\quad\ ${\Big\downarrow} \oplus N_b$};
	\node[above=0.02 of Eve, xshift=.7cm] (ze) {\quad\ ${\Big\downarrow} \oplus N_e$};
	
	\node[blank, left=.2 of Alice] (x) {$\underset{\in \Lambda_b/\Lambda_e}{X_m+X_r = X} \longrightarrow$};

	\node[blank, right=.1 of Bob] (yb) {$\leadsto\ Y_b = H_b X + N_b$};
	\node[blank, right=.1 of Eve] (ye) {$\leadsto\ Y_e = H_e X + N_e$};

	\draw[pil_rev] (Bob.west) -- node[pos=0.5, above]{\scriptsize $H_{b}$} (Alice.east);	
	\draw[pil] (Alice.east) |- node[pos=0.86, above]{\scriptsize $H_e$} (Eve.west);	
	\draw[black,thick,radiation,decoration={angle=45}] ([xshift=.5cm]Alice.north) -- +(45:0.5);

\end{tikzpicture}}
\caption{A wiretap communication setup. Alice communicates to Bob over a wiretap channel in the presence of Eve, the eavesdropper.}
\label{fig:wiretap}
\end{framed}
\end{figure}
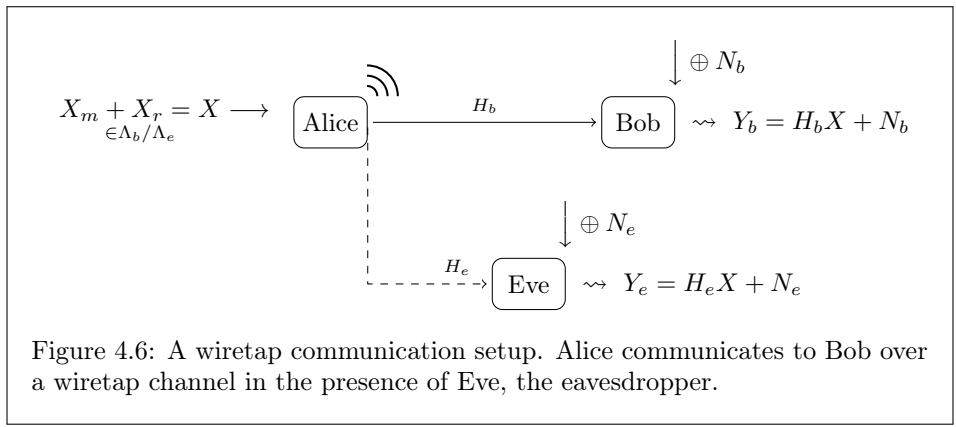

With this strategy, the lattice $\Lambda_b$ should be designed so that Bob can successfully decode the intended message, and it was shown for the SISO model in \cite{belfiore:wiretap} that the lattice $\Lambda_b$ needs to be designed as for regular fading channels. The MIMO setup was subsequently studied in \cite{belfiore:mimo_wiretap}. We illustrate this transmission model in Figure~\ref{fig:wiretap}. 

We are interested in the lattice $\Lambda_e$. We can interpret the channel equation as a Gaussian channel, where the transmitted signal is taken from a \emph{faded lattice} $\Lambda_H$ with generator matrix $M_{\Lambda_H} = (I_{T}\otimes \overline{H})M_{\Lambda}$ affected by the channel. We write $\Lambda_{b,H_b}$ and $\Lambda_{e,H_e}$ for the faded lattices related to Bob and Eve, respectively. 

In Publications VI and VII, our interest is in studying the design of the coarse lattice $\Lambda_e$ in the MIMO wiretap channel, given a fixed fine lattice $\Lambda_b$ that is good for Bob. In Publication VI, we adopt the usual probability theoretic approach to code design based on \emph{Eve's correct decoding probability} (ECDP), derived in \cite{belfiore:mimo_wiretap}. This quantity can be approximately upper bounded by an expression of the form 
\begin{align*}
	\mathrm{ECDP} \lesssim \sum\limits_{X \in \Lambda_e}{\det\left(I_n+\rho_e XX^\dagger\right)^{-(n_e+T)}},
\end{align*}
where $\rho_e$ denotes Eve's $\snr$.

In \cite{gnilke:siso_wiretap}, the authors show that well-rounded lattices offer good performance for the SISO wiretap channel. Based on the so-called first coding gain in the MIMO setup \cite{vehkalahti:min_delay}, 
\begin{align*}
	\delta_1(\Lambda_e) := \inf\left\{\left.\|X\|^2_F \right| X \in \Lambda_e\backslash\left\{0\right\}\right\},
\end{align*}
which determines the behavior of the employed lattice code in the low $\snr$ regime, we argue that the property of well-roundedness for the coarse lattice $\Lambda_e$ is also advantageous in the MIMO wiretap channel. To exemplify our findings, we fix the fine lattice $\Lambda_b$ to be the Alamouti or the Golden code, respectively, and compare the performance of several well-rounded and non-well-rounded sublattices of the two. The empirical results are unambiguous, showing that well-rounded lattices offer a formidable performance.

In Publication VII we consider the same problem but from an information theoretic perspective. Instead of the ECDP, our measure of performance is the mutual information between Eve's channel output and the original message, $\info\left[X; (Y_e,H_e)\right]$, which should be minimized. We consider two settings often found in literature.
\begin{itemize}
	\item[i)] The $\bmod\ \Lambda_s$ channel. Here, it is assumed that Eve only has knowledge of the equivalence class $Y_e/\Lambda_{s,H}$, where $\Lambda_s \subset \Lambda_e$ is a \emph{shaping lattice}. Alice chooses the random part $\mb{x}_r$ of the message uniformly at random from the finite set. 
	
	\item[ii)] The discrete Gaussian coset coding approach. Alice picks the random part of the message in such a way that the overall message $\mb{x} = \mb{x}_m + \mb{x}_r$ follows a lattice Gaussian distribution centered on the shifted lattice $\mb{x}_m + \Lambda_e$.  
\end{itemize}

For both setups, we derive variants of upper bounds on the mutual information \cite{mirghasemi:wiretap, luzzi:wiretap}. Our bounds are shown to be increasing functions of the expected flatness factor of the faded lattice related to the eavesdropper, $\mathbb{E}_H\left[\varepsilon_{\Lambda_{e,H_e}}(\sigma_e^2)\right]$, where $\sigma_e^2$ is the Eavesdropper's noise variance. Expectation is taken over all channel realizations. Independent of the setup, we conclude that the lattice $\Lambda_e$ should be designed so that its expected flatness factor is minimized. 

The result provided in Theorem~\ref{thm:theta_approx} then proves crucial to be able to provide an empirical analysis, as it renders the problem of computing the average flatness factor computationally inexpensive. In order to examine the predictive ability of the derived upper bounds in terms of actual performance, we compute the upper bound on the mutual information for the same lattices that are treated in \cite{gnilke:siso_wiretap}, where actual lengthy channel simulations are carried out. The obtained results are precisely what one would hope for, as there is a complete agreement between both approaches. In particular, well-rounded lattices again prove to be a particularly promising family of lattices.

\chapter{Conclusions}
\label{chp:conclusions}

The central topic of this thesis is the versatility of lattices in communications. Interesting mathematical objects themselves, lattices are indispensable in wireless communications and numerous performance criteria rely on the underlying lattice structure of a physical layer coding scheme. 

Given the rapid progress in wireless communications, it is not surprising that new communication protocols are proposed on a regular basis, such as the amplify-and-forward or compute-and-forward relaying protocols treated in this thesis. Older concepts, on the other hand, can suddenly become relevant in a wireless setting, for example wiretap coset coding.

In this thesis, we have considered multiple wireless communication settings and were particularly concerned with studying the design of the corresponding lattice codes. There are many aspects of a code that can be considered, be it improved reliability, reduced encoding and decoding complexity, reduced power consumption, increased rate, to name a few. In particular, developing efficient decoding algorithms that do not rely on suboptimal methods is an utterly important problem that often cannot be realized right away due to the complexity of the transmission protocols. 

All of the considered communication settings allow for many future research directions. A very challenging and interesting problem would be the adaptation of the compute-and-forward protocol to the MIMO setting. While this has already been attempted, the proposed adaptation suggests to code over space while avoiding spatial diversity. Thus, the development of an alike protocol which allows for the use of space--time codes and takes advantage of both types of diversity would certainly lead to interesting design questions for code construction in this scenario. 

Further, the work carried out in the contest of wiretap coset coding lead to the consideration of well-rounded lattices. These lattices are very interesting and useful, as for example the best sphere packing in any dimension is necessarily achieved by a well-rounded lattice. Yet, it is difficult to make general statements or even find families of well-rounded lattices in dimensions higher than two. Thus, the mathematical study of well-rounded lattices could prove to be very fruitful for applications in wireless communications. 

Lattices are not just the solution to a problem arising in wireless communications. More interestingly, many applications provide additional motivation for studying purely mathematical problems. It is this interplay of communications engineering and mathematics that allows a multidisciplinary thriving research community.

\renewcommand{\bibname}{Bibliography}
\bibliographystyle{plain}
\nocite{*}
\bibliography{references}

\begin{thebibliography}{10}

\bibitem{osi}
{OSI} model.
\newblock \url{https://en.wikipedia.org/wiki/OSI_model}.
\newblock Accessed 26.2.17.

\bibitem{alamouti:stc}
S.~Alamouti.
\newblock A simple transmitter diversity scheme for wireless communications.
\newblock {\em IEEE Journal on Selected Areas in Communications},
  16(8):1451--1458, 1998.

\bibitem{belfiore:lattices_caf}
J.-C. Belfiore.
\newblock Lattice codes for the compute-and-forward protocol: The flatness
  factor.
\newblock In {\em Proceedings of the IEEE Information Theory Workshop}, 2011.

\bibitem{belfiore:flatness_caf}
J.-C. Belfiore and C.~Ling.
\newblock The flatness factor in lattice network coding: Design criterion and
  decoding algorithm.
\newblock In {\em Zurich Seminar on Communications}, 2012.

\bibitem{belfiore:wiretap}
J.-C. Belfiore and F.~Oggier.
\newblock Lattice code design for the {R}ayleigh fading wiretap channel.
\newblock In {\em Proceedings of the IEEE International Conference on
  Communications Workshops}, 2011.

\bibitem{belfiore:mimo_wiretap}
J.-C. Belfiore and F.~Oggier.
\newblock An error probability approach to {MIMO} wiretap channels.
\newblock {\em IEEE Transactions on Communications}, 61(8):3396--3403, 2013.

\bibitem{belfiore:nvd}
J.-C. Belfiore and G.~Rekaya.
\newblock Quaternionic lattices for space--time coding.
\newblock In {\em Proceedings of the IEEE Information Theory Workshop}, 2003.

\bibitem{belfiore:golden}
J.-C. Belfiore, G.~Rekaya, and E.~Viterbo.
\newblock The golden code: a $2\times 2$ full-rate space--time code with
  non-vanishing determinants.
\newblock {\em IEEE Transactions on Information Theory}, 51(4):1432--1436,
  2005.

\bibitem{berhuy:fd}
G.~Berhuy, N.~Markin, and B.~A. Sethuraman.
\newblock Fast lattice decodability of space--time block codes.
\newblock In {\em Proceedings of the IEEE International Symposium on
  Information Theory}, 2014.

\bibitem{berhuy:fd_bounds}
G.~Berhuy, N.~Markin, and B.~A. Sethuraman.
\newblock Bounds on fast decodability of space--time block codes,
  skew-hermitian matrices, and {A}zumaya algebras.
\newblock {\em IEEE Transactions on Information Theory}, 61(4):1959--1970,
  2015.

\bibitem{biglieri:fd}
E.~Biglieri, Y.~Hong, and E.~Viterbo.
\newblock On fast-decodable space--time block codes.
\newblock {\em IEEE Transactions on Information Theory}, 55(2):524--530, 2009.

\bibitem{conway:voronoi}
J.~H. Conway and N.~J.~A. Sloane.
\newblock A fast encoding method for lattice codes and quantizers.
\newblock {\em IEEE Transactions on Information Theory}, 29(6):820--824, 1983.

\bibitem{conway:lattices}
J.~H. Conway and N.~J.~A. Sloane.
\newblock {\em Sphere Packings, Lattices and Groups}.
\newblock Springer-Verlag, third edition, 1999.

\bibitem{ebeling:lattices}
W.~Ebeling.
\newblock {\em Lattices and Codes}.
\newblock Spektrum Verlag, third edition, 2013.

\bibitem{elia:perfect}
P.~Elia, B.~A. Sethuraman, and P.~V. Kumar.
\newblock Perfect space--time codes for any number of antennas.
\newblock {\em IEEE Transactions on Information Theory}, 53(11):3853--3868,
  2007.

\bibitem{fukshansky:bounds_frobenius}
L.~Fukshansky and A.~Schürmann.
\newblock Bounds on generalized frobenius numbers.
\newblock {\em European Journal of Combinatorics}, 42(3):361--368, 2011.

\bibitem{gnilke:siso_wiretap}
O.~Gnilke, H.~Tran, A.~Karrila, and C.~Hollanti.
\newblock Well-rounded lattices for reliability and security in {R}ayleigh
  fading {SISO} channels.
\newblock In {\em Proceedings of the IEEE Information Theory Workshop}, 2016.

\bibitem{hollanti:thesis}
C.~Hollanti.
\newblock {\em Order-Theoretic Methods for Space--Time Coding: Symmetric and
  Asymmetric Designs}.
\newblock PhD thesis, University of Turku, 2009.

\bibitem{hollanti:asymmetric}
C.~Hollanti and H.~f.~Lu.
\newblock Construction methods for asymmetric and multiblock space--time codes.
\newblock {\em IEEE Transactions on Information Theory}, 55(3):1086--1103,
  2009.

\bibitem{hollanti:order1}
C.~Hollanti and J.~Lahtonen.
\newblock A new tool: Constructing {STBC}s from maximal orders in central
  simple algebras.
\newblock In {\em Proceedings of the IEEE Information Theory Workshop}, 2006.

\bibitem{hollanti:relay1}
C.~Hollanti and N.~Markin.
\newblock Algebraic fast-decodable relay codes for distributed communications.
\newblock In {\em Proceedings of the IEEE International Symposium on
  Information Theory}, 2012.

\bibitem{hollanti:relay2}
C.~Hollanti and N.~Markin.
\newblock A unified framework for constructing fast-decodable codes for {N}
  relays.
\newblock In {\em Proceedings of the 20th International Symposium on
  Mathematical Theory of Networks and Systems}, 2012.

\bibitem{jafarkhani:stc}
H.~Jafarkhani.
\newblock {\em Space--Time Coding: Theory and Practice}.
\newblock Cambridge University Press, 2005.

\bibitem{jithamithra:sphere_decod}
G.~R. Jithamithra and B.~S. Rajan.
\newblock Minimizing the complexity of fast sphere decoding of {STBC}s.
\newblock {\em IEEE Transactions on Wireless Communications},
  12(12):6142--6153, 2013.

\bibitem{luzzi:wiretap}
L.~Luzzi, C.~Ling, and R.~Vehkalahti.
\newblock Almost universal codes for fading wiretap channels.
\newblock In {\em Proceedings of the IEEE International Symposium on
  Information Theory}, 2016.

\bibitem{markin:iterated}
N.~Markin and F.~Oggier.
\newblock Iterated space--time code constructions from cyclic algebras.
\newblock {\em IEEE Transactions on Information Theory}, 59(9):5966--5979,
  2013.

\bibitem{mejri:fd_revisited}
A.~Mejri, M.-A. Khsiba, and G.~Rekaya-Ben Othman.
\newblock Reduced-complexity {ML} decodable {STBC}s: Revisited design criteria.
\newblock In {\em Proceedings of the International Symposium on Wireless
  Communication Systems}, 2015.

\bibitem{mejri:decoding_caf}
A.~Mejri and G.~Rekaya-Ben Othman.
\newblock Efficient decoding algorithms for the compute-and-forward strategy.
\newblock {\em IEEE Transactions on Communications}, 63(7):2475--2485, 2015.

\bibitem{milne:cft}
J.~Milne.
\newblock Class field theory.
\newblock \url{http://jmilne.org/math/CourseNotes/CFT.pdf}, 2013.
\newblock Graduate course notes, v4.02.

\bibitem{milne:ant}
J.~Milne.
\newblock Algebraic number theory.
\newblock \url{http://jmilne.org/math/CourseNotes/ANT.pdf}, 2014.
\newblock Graduate course notes, v2.0.

\bibitem{mirghasemi:wiretap}
H.~Mirghasemi and J.-C. Belfiore.
\newblock Lattice code design criterion for {MIMO} wiretap channels.
\newblock In {\em Proceedings of the IEEE Information Theory Workshop}, 2015.

\bibitem{nabar:relay}
R.~U. Nabar, H.~Bölcskei, and F.~W. Kneubühler.
\newblock Fading relay channels: performance limits and space--time signal
  design.
\newblock {\em IEEE Journal on Selected Areas in Communications},
  22(6):1099--1109, 2004.

\bibitem{nazer:cf}
B.~Nazer and M.~Gastpar.
\newblock Compute-and-forward: Harnessing interference through structured
  codes.
\newblock {\em IEEE Transactions on Information Theory}, 57(10):6463--6486,
  2011.

\bibitem{neukirch:ant}
J.~Neukirch.
\newblock {\em Algebraic Number Theory}.
\newblock Springer-Verlag, 2010.

\bibitem{oggier:perfect}
F.~Oggier, G.~Rekaya, J.-C. Belfiore, and E.~Viterbo.
\newblock Perfect space--time block codes.
\newblock {\em IEEE Transactions on Information Theory}, 52(9):3885--3902,
  2006.

\bibitem{osmane:caf_implementation}
A.~Osmane and J.-C. Belfiore.
\newblock The compute-and-forward protocol: Implementation and practical
  aspects.
\newblock arXiv:1107.0300v1, 2011.

\bibitem{ozarov:wiretap}
L.~H. Ozarov and A.~D. Wyner.
\newblock Wire-tap channel {II}.
\newblock {\em AT\&T Bell Laboratories Technical Journal}, 63(10):2135--2157,
  1984.

\bibitem{paredes:silver}
J.~Paredes, A.~B. Gershman, and M.~G. Alkhanari.
\newblock A $2\times 2$ space--time code with non-vanishing determinants and
  fast maximum likelihood decoding.
\newblock In {\em Proceedings of the IEEE International Conference on
  Acoustics, Speech, and Signal Processing}, 2007.

\bibitem{poltyrev:nested}
G.~Poltyrev.
\newblock On coding without restrictions for the {AWGN} channel.
\newblock {\em IEEE Transactions on Information Theory}, 40(2):409--417, 1994.

\bibitem{rekaya:nvd}
G.~Rekaya, J.-C. Belfiore, and E.~Viterbo.
\newblock Algebraic $3\times 3$, $4\times 4$ and $6\times 6$ space--time codes
  with non-vanishing determinants.
\newblock In {\em Proceedings of the IEEE International Symposium on
  Information Theory and its Applications}, 2004.

\bibitem{sethuraman:stc}
B.~A. Sethuraman, B.~S. Rajan, and V.~Shashidhar.
\newblock Full-diversity, high-rate space--time block codes from division
  algebras.
\newblock {\em IEEE Transactions on Information Theory}, 49(10):2596--2616,
  2003.

\bibitem{tarokh:stc}
V.~Tarokh, N.~Seshadri, and A.~R. Calderbank.
\newblock Space--time codes for high data rate wireless communication:
  Performance criterion and code construction.
\newblock {\em IEEE Transactions on Information Theory}, 44(2):744--765, 1998.

\bibitem{unger:quaternion}
T.~Unger and N.~Markin.
\newblock Quadratic forms and space--time block codes from generalized
  quaternion and biquaternion algebras.
\newblock {\em IEEE Transactions on Information Theory}, 57(9):6148--6156,
  2011.

\bibitem{vehkalahti:min_delay}
R.~Vehkalahti and C.~Hollanti.
\newblock Reducing complexity with less than minimum delay space--time lattice
  codes.
\newblock In {\em Proceedings of the IEEE Information Theory Workshop}, 2011.

\bibitem{vehkalahti:dense_mimo}
R.~Vehkalahti, C.~Hollanti, J.~Lahtonen, and K.~Ranto.
\newblock On the densest {MIMO} lattices from cyclic division algebras.
\newblock {\em IEEE Transactions on Information Theory}, 55(8):3751--3780,
  2009.

\bibitem{vehkalahti:asymmetric}
R.~Vehkalahti, C.~Hollanti, and F.~Oggier.
\newblock Fast-decodable asymmetric space--time codes from division algebras.
\newblock {\em IEEE Transactions on Information Theory}, 58(4):2362--2385,
  2012.

\bibitem{widmer:counting_lattice_points}
M.~Widmer.
\newblock Lipschitz class, narrow class, and counting lattice points.
\newblock {\em Proceedings of the American Mathematical Society},
  140(2):677--689, 2011.

\bibitem{wyner:wiretap}
A.~D. Wyner.
\newblock The wire-tap channel.
\newblock {\em Bell System Technical Journal}, 54(8):1355–1387, 1975.

\bibitem{yang:af}
S.~Yang and J.-C. Belfiore.
\newblock Optimal space--time codes for the {MIMO} amplify-and-forward
  cooperative channel.
\newblock {\em IEEE Transactions on Information Theory}, 53(2):647--663, 2007.

\bibitem{zamir:lattices}
R.~Zamir.
\newblock Lattices are everywhere.
\newblock In {\em Proceedings of the IEEE Information Theory and Applications
  Workshop}, 2009.

\bibitem{zamir:nested}
R.~Zamir and M.~Feder.
\newblock On lattice quantization noise.
\newblock {\em IEEE Transactions on Information Theory}, 42(4):1152--1159,
  1996.

\end{thebibliography}

\end{document}